%% file: DistortedHKModel.tex
\documentclass[letterpaper,superscriptaddress,english,aps,prb,twocolumn,floatfix, showpacs, amsfonts, amssymb]{revtex4}
\usepackage{epsfig,graphicx,psfrag,amsmath,amssymb,float}
\usepackage[T1]{fontenc}
\usepackage[latin9]{inputenc}
\usepackage{amsmath}
\usepackage{amssymb}

\usepackage{amscd}
\usepackage{bm}
\usepackage{psfrag}

\usepackage{babel}
\usepackage{times}

\usepackage{hyperref}
\hypersetup{colorlinks=true}

\newcommand{\be}{\begin{equation}}
\newcommand{\ee}{\end{equation}}
\newcommand{\bea}{\begin{eqnarray}}
\newcommand{\eea}{\end{eqnarray}}

\newcommand{\vmN}{\ensuremath{{\bf m}_{\text N}}}
\newcommand{\vmS}{\ensuremath{{\bf m}_{\text S}}}
\newcommand{\mN}{\ensuremath{m_{\text N}}}
\newcommand{\mS}{\ensuremath{m_{\text S}}}
\newcommand{\QN}{\ensuremath{Q_2^{\text N}}}
\newcommand{\QS}{\ensuremath{Q_2^{\text S}}}
\newcommand{\Na}{Na$_2$IrO$_3$}
\newcommand{\Li}{Li$_2$IrO$_3$}

\DeclareMathOperator{\tr}{tr}

\begin{document}

\title{Order-by-disorder  and spin-orbital liquids in a distorted Heisenberg-Kitaev model}

\author{Eran Sela}
\affiliation{Raymond and Beverly Sackler School of Physics and Astronomy, Tel-Aviv University, Tel Aviv, 69978, Israel}

\author{Hong-Chen Jiang}
\affiliation{Department of Physics, University of California, Berkeley, California 94720, USA}

\author{Max H. Gerlach}
\affiliation{Institute for Theoretical Physics, University of Cologne, 50937 Cologne, Germany}

\author{Simon Trebst}
\affiliation{Institute for Theoretical Physics, University of Cologne, 50937 Cologne, Germany}

\date{\today}

\begin{abstract}
The microscopic modeling of  spin-orbit entangled $j=1/2$ Mott insulators such as the layered hexagonal Iridates \Na\  and \Li\  has spurred an interest in the physics of Heisenberg-Kitaev models. Here we explore the effect of lattice distortions on the formation of the collective spin-orbital states which include not only conventionally ordered phases but also gapped and gapless spin-orbital liquids. In particular, we demonstrate that in the presence of spatial anisotropies of the exchange couplings  conventionally ordered states are formed through an order-by-disorder selection which is not only sensitive to the type of exchange anisotropy but also to the relative strength of the Heisenberg and Kitaev couplings.
The spin-orbital liquid phases of the Kitaev limit -- a gapless phase in the vicinity of spatially isotropic couplings and a gapped Z$_2$ phase for a dominant spatial anisotropy of the exchange couplings --  show vastly different sensitivities to the inclusion of a Heisenberg exchange. While the gapless phase is remarkably stable, the gapped Z$_2$ phase quickly breaks down in what might be a rather unconventional phase transition driven by the simultaneous condensation of its elementary excitations.
\end{abstract}


\maketitle

\section{Introduction}
\begin{figure*}[t]
  \begin{center}
    \includegraphics*[width=\linewidth]{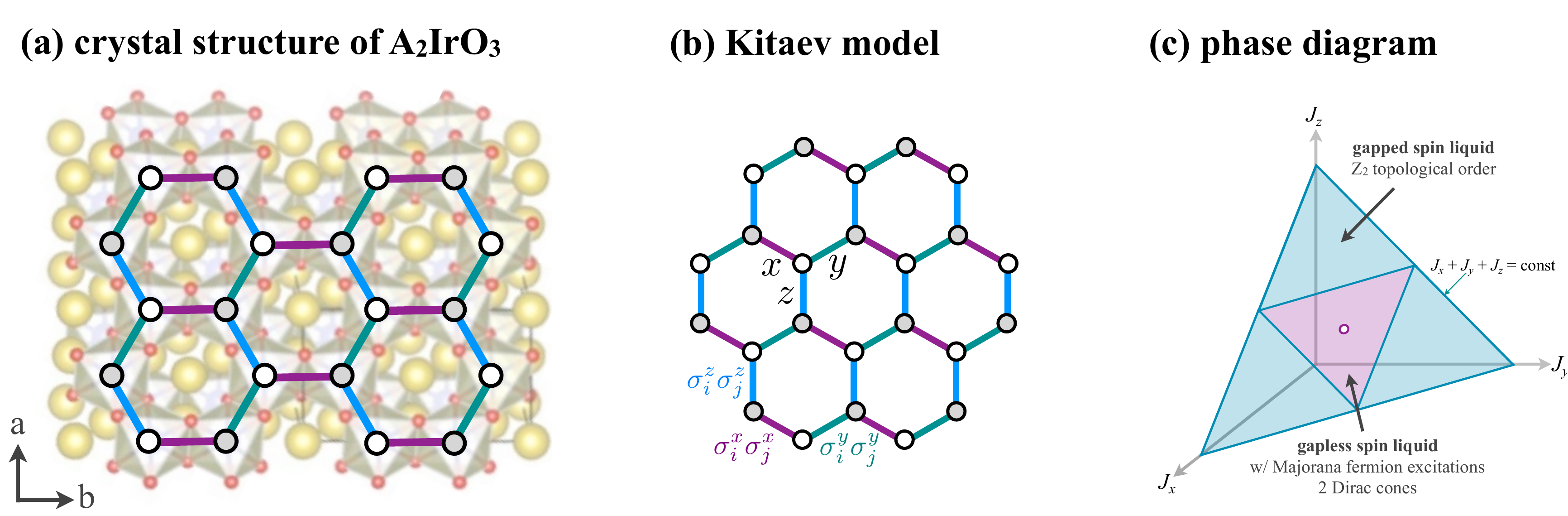}
    \caption{
        \label{fig:kitaev}
        (a) Crystal structure of the layered Iridates A$_2$IrO$_3$ with A\,=\,(Na, Li).
        (b) Sketch of the microscopic interactions in the Kitaev honeycomb model.
        (c) Phase diagram of the quantum Kitaev model.
    }
  \end{center}
\end{figure*}

The intricate interplay of electronic correlations, spin-orbit coupling, and crystal-field effects in $5d$ transition metal oxides
has led to the discovery of an intriguing variety of quantum states of matter including  Weyl semi-metals, axion insulators,
or topological Mott insulators
\cite{Witczak-Krempa14}.
In the correlation dominated regime unusual local moments such as spin-orbit entangled degrees of freedom can form
and whose collective behavior gives rise to unconventional types of magnetism including the formation of quadrupolar correlations
or the emergence of so-called spin liquid states.\cite{Balents10}
On the materials side a particularly prolific group of compounds are the Iridates, whose electronic state can be either weakly
conducting or insulating. Common to all Iridates is that the Iridium ions typically occur in an Ir$^{4+}$ ionization state corresponding to a $5d^5$ electronic configuration. For the insulating compounds a particularly intriguing scenario is the formation of a so-called $j=1/2$ Mott insulator \cite{kim08,kim09}, in which a crystal field splitting of the $d$ orbitals into t$_{2g}$ and e$_g$ orbitals and a subsequent spin-orbit entanglement leads to a Mott transition yielding a completely filled $j=3/2$ state and a half-filled $j=1/2$ doublet.
The microscopic exchange between these spin-orbit entangled $j=1/2$ local moments has been argued \cite{jackeli,chaloupka} to give rise to  interactions which combine a spin-like contribution in form of an isotropic Heisenberg exchange with an orbital-like contribution in form of a highly anisotropic exchange whose easy axis depends on the spatial orientation of the exchange path. Such orbital exchange interactions are well known from the early work of Kugel and Khomskii \cite{Khomskii} on quantum compass models \cite{Nussinov13}
to induce a high level of exchange frustration, i.e. they inhibit an ordering transition of the local moments which cannot simultaneously align with all their nearest neighbors due to the competing orientations of the respective easy axis. This frustration mechanism is particularly effective in the so-called Kitaev model \cite{Kitaev}, a honeycomb compass model where the exchange easy axis points along the $x$, $y$, and $z$-directions for the three different bond orientations in the honeycomb lattice, see Fig.~\ref{fig:kitaev}(b). Its phase diagram parametrized in the relative coupling strength of the three types of exchanges exhibits two incarnations of spin liquid phases: an extended gapless spin liquid phase around the point of equally strong exchange interactions and gapped $Z_2$ spin liquid phases if one of the three coupling strengths dominates, see Fig.~\ref{fig:kitaev}(c) for a detailed phase diagram.
On the materials side, the layered Iridates Na$_2$IrO$_3$ and Li$_2$IrO$_3$, which form $j=1/2$ Mott insulators with the Iridium ions arranged on a hexagonal lattice as illustrated in Fig.~\ref{fig:kitaev}(a), have recently attracted considerable attention as possible solid state incarnations \cite{jackeli,chaloupka,singh10,singh12,choi,ye} of the Heisenberg-Kitaev model 
\footnote{Generalizations of the Heisenberg-Kitaev model to lattice geometries beyond the hexagonal lattice have recently been considered in both two and three spatial dimensions \cite{Daghofer12,Kimchi14,Kimchi13,Kim14,Nasu14,Lee14a,Lee14b,Hermanns14} motivated in part by the recent synthesis of three-dimensional honeycomb Iridates \cite{Analytis14,Takagi14}.}. 

In this manuscript, we inspect the role of distortions, i.e. spatial anisotropies of the exchange coupling strength, on the collective spin-orbital state of the hexagonal Heisenberg-Kitaev model away from the exactly solvable Kitaev limit. Our motivation to do so has been twofold.
First, early space group determinations of the layered Iridate \Na\  using powder x-ray diffraction scans \cite{singh10} hinted at space group C2/{\em c}, in which the hexagonal lattice formed by the Ir$^{4+}$ ions is slightly distorted along one of its three principal directions. However, more refined inelastic neutron scattering \cite{choi} and single-crystal x-ray diffraction measurements \cite{ye} later revealed that the correct space group of \Na\  is in fact space group C2/{\em m} and the hexagonal lattice formed by the Ir$^{4+}$ ions is an almost perfectly 120$^\circ$ symmetric honeycomb lattice. As we will show in this manuscript the collective spin-orbital states of these systems are nevertheless highly sensitive to small spatial anisotropies of the exchange couplings, which experimentally can be probed via external pressure measurements inducing small lattice distortions and concurrent exchange anisotropies.
Second, we hoped to shed further light on the putative quantum critical point in the undistorted Heisenberg-Kitaev model~\cite{chaloupka,jiang,schaffer,reuther} between a gapless spin-orbital liquid phase extending out of the Kitaev limit and a conventionally ordered ``stripy'' phase for the intermediate regime of roughly equally strong Heisenberg and Kitaev couplings. Our analysis shows that exchange coupling distortions are relevant perturbations in any field theoretical description of such a quantum critical point, which depending on their relative strength induce different types of conventionally ordered states in an order-by-disorder selection. This mechanism, which for an infinitesimally small distortion selects a subset of the six possible stripy spin-orbital orderings of the undistorted model, is at play for the entire stripy phase of the  Heisenberg-Kitaev model in the intermediate coupling regime. In fact, the selection process turns out to be subtly sensitive not only on the sign of the distortion but also the relative coupling strength of Heisenberg and Kitaev exchange which leads to a total of four different stripy ordered phases in the phase diagram of the distorted Heisenberg-Kitaev model.

We will start our discussion by first considering the classical variant of the distorted Heisenberg-Kitaev model in section \ref{sec:classical}. The phase diagram of the classical model already includes all of the conventionally ordered phases found in its quantum mechanical counterpart as well as its own variation of an order-by-disorder selection of ordered states in the presence of exchange coupling distortions.
The entire phase diagram of the classical model as well as its finite-temperature behavior are discussed via extensive numerical simulations.
We further consider in detail the classical limit of the Kitaev model, which in the absence of distortions is known to exhibit a classical spin liquid state with Coulomb gas correlations.\cite{chandra}  We show that the inclusion of exchange distortions leads to a break-down of these power-law correlations and a partial lifting of the residual entropy at zero-temperature, which is also reflected in characteristic signatures of the low-temperature specific heat behavior.
We then turn to the quantum Heisenberg-Kitaev model in section \ref{sec:quantum} whose phase diagram we have determined via extensive numerical simulations relying on the density matrix renormalization group (DMRG) on finite two-dimensional clusters. The quantum order-by-disorder selection is discussed and found to be in perfect agreement with the numerical data. Finally we discuss the possibility of an exotic continuous quantum phase transition, where the Heisenberg exchange drives the system out of the gapped $Z_2$ spin liquid phase of the distorted Kitaev model into a stripy ordered phase. Based on perturbative arguments we conjecture that this transition might be driven by the simultaneous condensation of the excitations of the $Z_2$ spin liquid.
We round off the manuscript with a summary and outlook in section \ref{sec:outlook}.

%
%

\section{Classical Heisenberg-Kitaev model}
\label{sec:classical}

\begin{figure*}
  \includegraphics[width=\linewidth]{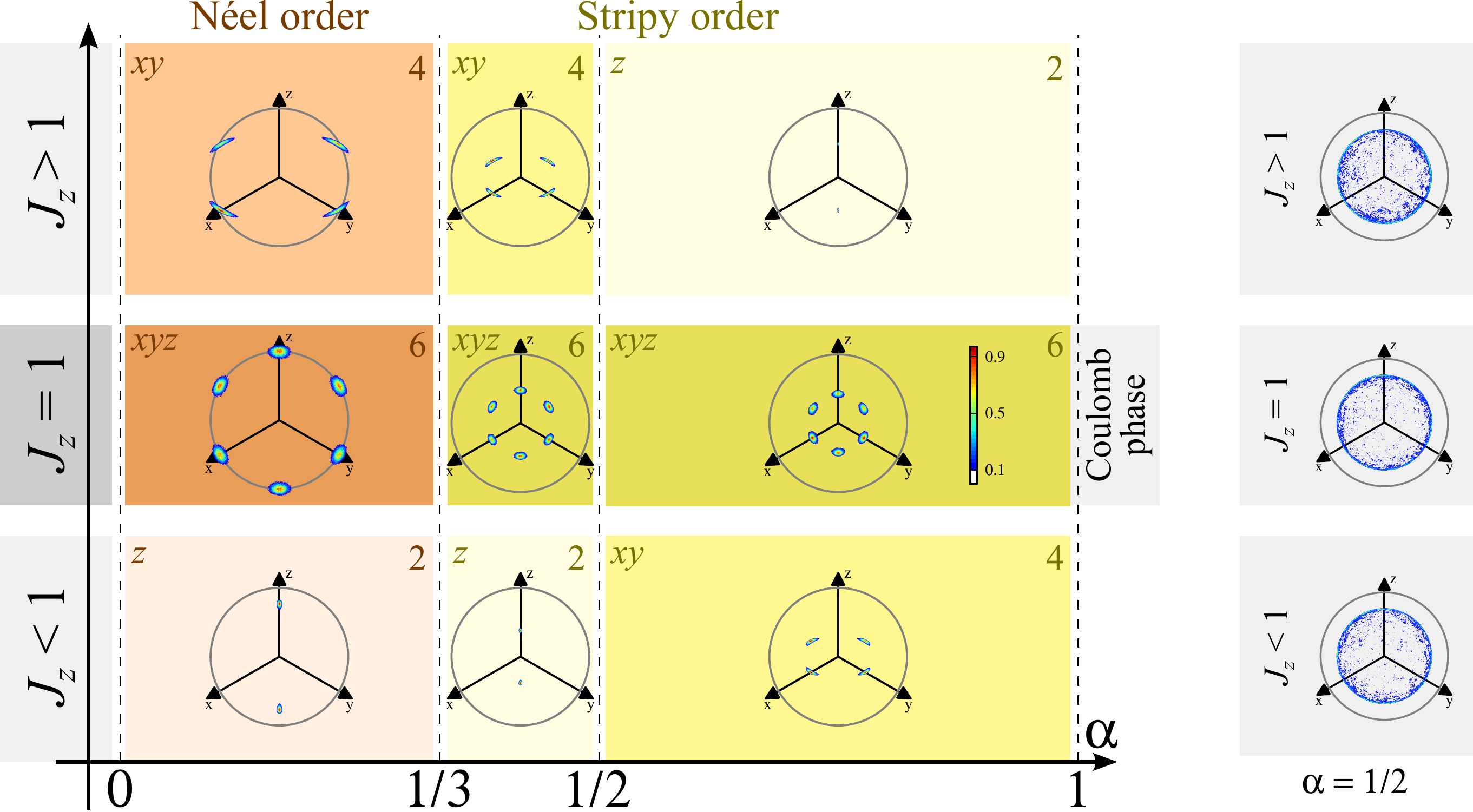}
  \caption{\label{fig:distortedphasediagram} Low-temperature phase     diagram of the Heisenberg-Kitaev model under variation of the     relative strength $\alpha$ of the Heisenberg and Kitaev couplings     and the distortion $J_z$.  For $0 < \alpha < 1/3$ we additionally     show histograms of the N\'eel magnetization $\vmN$, while the     histograms for $1/3 < \alpha < 1/2$ and $1/2 < \alpha < 1$ display     the stripy order parameter $\vmS$.  In the undistorted case with     $J_z=1$ the magnetization vector lies on one of the cubic axes,     either in positive or negative direction, yielding a sixfold     degeneracy.  In the distorted models with $J_z \gtrless 1$     depending on $\alpha$ there is either a twofold degeneracy with     the magnetization pointing in $\pm z$-direction or a fourfold     degeneracy where the magnetization points in one of the $\pm x$-     or $\pm y$-directions.  The color coding is according to a     normalization by highest density.  Each histogram has been     measured in a single parallel-tempering simulation of a system of     size $L=32$.  At $\alpha=1/2$ the model is $O(3)$ symmetric and     hence we find no preferred directions of ordering in the     $\vmS$-histograms (right hand side).}
\end{figure*}

We start our discussion of the distorted Heisenberg-Kitaev model by first considering its classical version. Its Hamiltonian is given by
\begin{eqnarray}
   H & = &  (1-\alpha) \mathcal{H}_{\rm Heisenberg} - 2\alpha \mathcal{H}_{\rm Kitaev} \nonumber \\
      & = &  \sum_{\langle ij \rangle, \gamma} J_\gamma\, \left( (1-\alpha)\, {\bf S}_i {\bf S}_{j}  - 2\alpha\,  S^\gamma_i S^\gamma_{j} \right) \,,
      \label{eq:model}
\end{eqnarray}
where the spins ${\bf S}$ are classical $O(3)$ Heisenberg spins and the sums run over nearest neighbor bonds $\langle ij \rangle$ along
the three principal directions $\gamma$ of the honeycomb lattice labeled $x$,$y$, and $z$, see Fig.~\ref{fig:kitaev}(b). The coupling constants $0 < J_\gamma$ parametrize the overall strength of the couplings along these three bonds, while the parameter $0 < \alpha < 1$ parametrizes the relative strength of the Heisenberg and Kitaev exchange with $\alpha=0$ corresponding to the Heisenberg limit and $\alpha=1$ corresponding to the Kitaev limit. Note that the Heisenberg exchange is always antiferromagnetic, while the Kitaev exchange is always ferromagnetic. The choice of these coupling signs is motivated by the microscopic modeling \cite{chaloupka} of the layered Iridate compounds \Na\  and \Li.
To be even more explicit, the Hamiltonian can be decomposed into three types of bond terms which read
 \bea
 \label{eq:bondH}
H^x_{ij} & = & J_x[(1-\alpha) {\bf S}_i {\bf S}_{j} - (2 \alpha) S^x_i S^x_{j}], \nonumber \\
H^y_{ij} & = & J_y[(1-\alpha) {\bf S}_i {\bf S}_{j} - (2 \alpha) S^y_i S^y_{j}], \nonumber \\
H^z_{ij} & = & J_z[(1-\alpha) {\bf S}_i {\bf S}_{j} - (2 \alpha) S^z_i S^z_{j}].
\eea
The case of $J_x = J_y = J_z$ corresponds to spatially isotropic coupling strengths and the model reflects the $C_3$ rotational symmetry of the honeycomb lattice. We refer to this case as the undistorted Heisenberg-Kitaev model. To consider the effect of distortions, i.e. spatially anisotropic coupling strengths, we will vary the relative strength of the $J_z$ bond exchange while keeping the other two coupling strengths equal, i.e. $J_x = J_y$. We further use the convention that the overall coupling strength is constant, i.e. $J_x + J_y + J_z =3$, so that for varying $0 < J_z < 3$ we have  $J_x = J_y = (3 - J_z)/2$.


\subsection{Phase diagram of the distorted HK model}
\label{pdclassical}

A summary of the low-temperature ordered states of this classical model is provided in the phase diagram of Fig.~\ref{fig:distortedphasediagram}. The model exhibits a number of conventionally ordered states which we will discuss
in the following.

We start by surveying the phases of the undistorted, $C_3$ symmetric model for $J_z=1$,  see the center row of
Fig.~\ref{fig:distortedphasediagram}. At $\alpha=0$ we have an antiferromagnetic Heisenberg interaction stabilizing
a N\'eel ordered phase with a staggered moment pointing along an arbitrary direction.
Including a small (ferromagnetic) Kitaev interaction lowers the continuous O(3) symmetry of the Heisenberg model to a set of discrete symmetries including (i) time reversal symmetry, (ii) a $2\pi/3$ spin rotation about the [111] spin axis along with $C_3$ lattice rotations about an arbitrary site, and (iii) an inversion symmetry around any plaquette or bond center. Yet the N\'eel order survives. Interestingly, the direction of the N\'eel staggered moment is determined by a classical order-by-disorder mechanism, which we will discuss in more detail in Section~\ref{se:orderbydisorder}. Upon further increasing the Kitaev exchange the system will eventually disfavor N\'eel order and undergo a first-order transition to an alternate ordered state exhibiting ``stripy'' order. To see the order of the resulting phase, fortunately, at $\alpha=1/2$ after an appropriate change of spin variables the Hamiltonian reduces again to an O(3) symmetric model, albeit a ferromagnetic one~\cite{chaloupka}.

\begin{figure}[b]
  \begin{center}
    \includegraphics[width=0.8\columnwidth]{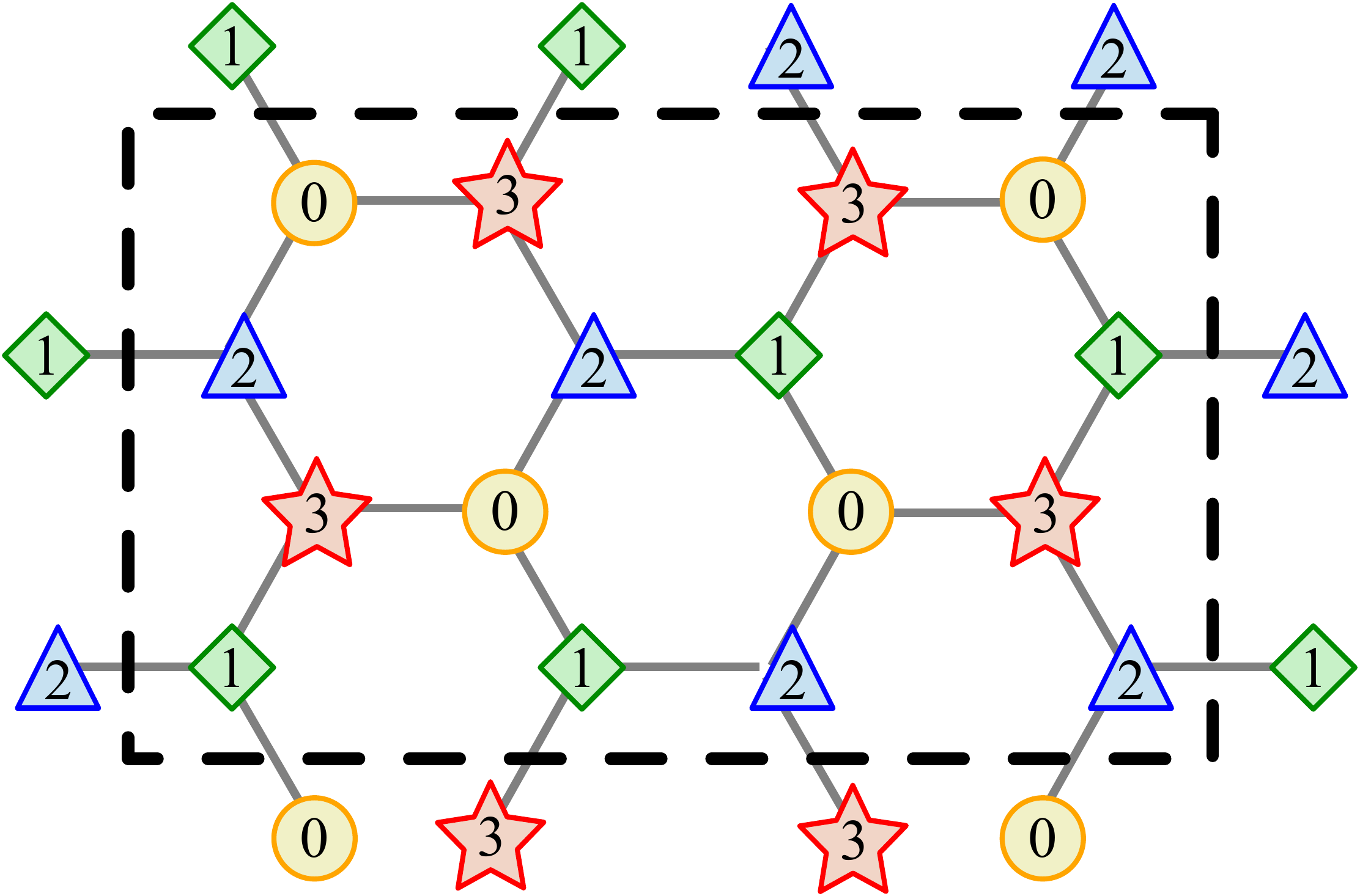}
    \caption{
        \label{fg:supercell}
        16 site supercell used to transform the Hamiltonian at         $\alpha=1/2$ to a O(3) symmetric ferromagnetic one.  Note that         we use a smaller unit cell compared to Ref.~\ref{CJK}.  }
  \end{center}
\end{figure}

We briefly describe the four-sublattice basis transformation. Note that at $\alpha=1/2$ the spin-spin interactions between $x$, $y$ and $z$ spin components have equal magnitude but depending on the bond type two interactions are antiferromagnetic and one is ferromagnetic. This interaction  can be transformed to a fully ferromagnetic one upon a relative $\pi$-rotation of the two spins around the special axis. We denote the new spin variables by ${\tilde{\bf S}}$. Explicitly, to make this transformation on the full lattice we define a 16 site supercell with sites of types $0,1,2,3$ as depicted in Fig.~\ref{fg:supercell}. The new spin variables  $\tilde{\bf S}$ are obtained by a $\pi$-rotation around $x$, $y$, or $z$ for sites of type 1, 2, and 3, respectively, and they are simply equal to ${\bf S}$ on sites of type 0.

\begin{figure}[t]
  \begin{center}
    \includegraphics[width=\linewidth]{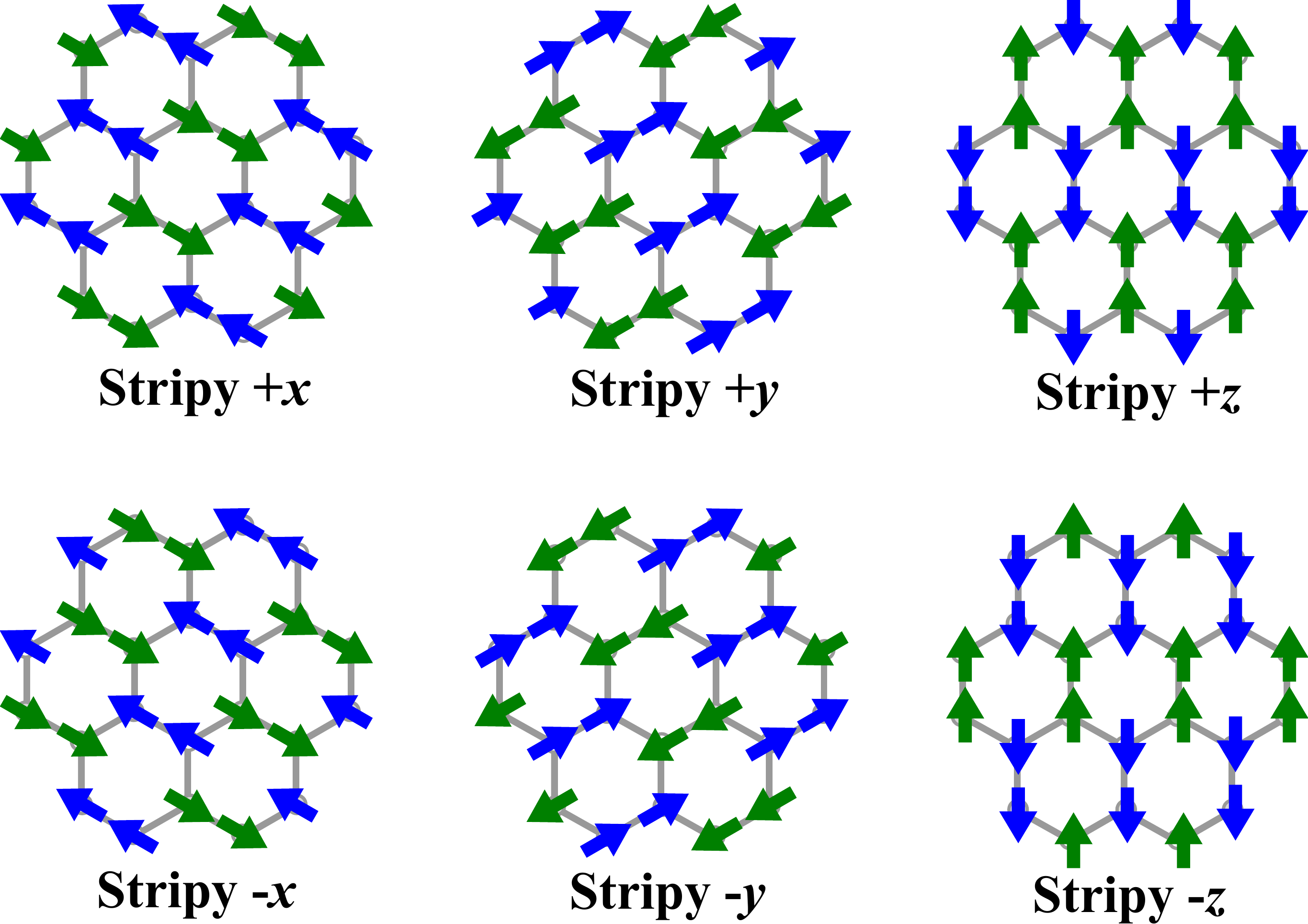}
    \caption{\label{fg:stripy} Illustration of the stripy $x$, $y$ and       $z$ phases where the arrows indicate the spin alignment along       the $x$, $y$, and $z$ spin directions. }
  \end{center}
\end{figure}

After the four-sublattice basis transformation the Hamiltonian in the new spin variables reads~\cite{chaloupka}
\bea
   \label{CJK}
H^\gamma_{ij} = J_\gamma[(\alpha-1) {\tilde{\bf S}}_i {\tilde{\bf S}}_{j} +2  (1-2 \alpha) \tilde{S}^\gamma_i \tilde{S}^\gamma_{j}].
\eea
Thus we see that at $\alpha=1/2$ the system has O(3) symmetry. The ground state is a ferromagnet in the $\tilde{S}$ variables. This translates to the stripy phases of the original spins; see Fig.~\ref{fg:stripy}. Similar to the Heisenberg point at $\alpha=0$, also at $\alpha=1/2$ the direction of the ferromagnetic moment is arbitrary due to the O(3) symmetry. But any finite deviation from $\alpha=1/2$ breaks the continuous symmetry down to a discrete one and we expect the ferromagnetic magnetization direction to be fixed at one of few discrete possibilities. As will be seen in section~\ref{se:orderbydisorder} this happens by a classical order by disorder mechanism.
The N\'eel and stripy phases have direct analogs in the quantum case. The most interesting quantum phase occurring for $\alpha \to 1$, which is a spin liquid with gapless excitations in the form of emergent Majorana fermions, does not have an immediate classical analog. Instead,  the system forms a classical spin liquid state -- a so-called Coulomb gas,~\cite{chandra} which exists only in the Kitaev limit, i.e. $\alpha = 1$, to which we will devote special attention in section~\ref{se:Coulomb}.

We now consider a finite amount of distortion $J_z \ne 1$. $J_z>1$ corresponds to strong dimers, while $J_z<1$ corresponds to dominating chains.
As can be easily obtained by calculating the energies of the various ordered states discussed, the N\'eel ordered region splits up into one ($J_z>1$) in which spins are in the $xy$ plane and another one ($J_z<1$) at which they point along the $z$ direction. Also in the stripy phases spins either point along $z$ for $(\alpha-1/2)(J_z-1)>0$, or they lie in the $xy$ plane for $(\alpha-1/2)(J_z-1)<0$. Note, that from pure energetics the directions of the spins in the $xy$ plane is not fixed. Also here the finite temperature order by disorder mechanism come to play; see section~\ref{se:orderbydisorder}.

The paragraph above relies on the following expressions for the energy per unit cell of the N\'eel and stripy phases with spins pointing along $\gamma$,
\bea
\label{meanfieldenergy}
E_{\text{N\'eel}\,\gamma}  &=&-  J_{\gamma} (1-3\alpha) - \sum_{\gamma' \ne \gamma} J_{\gamma'} (1-\alpha), \nonumber \\
E_{\text{stripy}\,\gamma} &=&J_{\gamma} (1-3\alpha)+ \sum_{\gamma' \ne \gamma} J_{\gamma'}
(\alpha-1).  \eea Also the N\'eel-stripy phase
transition lines can be found by equating energies. From \bea
E_{\text{N\'eel}\,z} =E_{\text{stripy}\,z}, \eea we obtain $\alpha =
1/3$, giving the line boundary between N\'eel and stripy for
$J_z<1$. By comparing \bea E_{\text{N\'eel}\,xy}
=E_{\text{stripy}\,xy}, \eea we also obtain $\alpha = 1/3$, giving the
line boundary between N\'eel and stripy phases for $J_z>1$. As a
result there is a straight vertical line at $\alpha=1/3$ marking the
N\'eel-stripy transition in the low-temperature phase diagram of
Fig.~\ref{fig:distortedphasediagram}.

As outlined in section~\ref{sec:numerical-results} we numerically obtain
the finite-temperature phase diagram shown in
Fig.~\ref{fig:transitiontemperatures}.


\subsection{Order by disorder and effective Ginzburg-Landau theory}
\label{se:orderbydisorder}
At $\alpha=1/2$ the magnetization points along an arbitrary direction due to the O(3) symmetry explicitly apparent in Eq.~(\ref{CJK}) (we refer to the $\tilde{\bf S}$ variables in terms of which the Hamiltonian is ferromagnetic). At finite deviations from this symmetric point one expects the Kitaev anisotropic interactions to stabilize a discrete set of orientations of the magnetization. However, as Eq.~(\ref{meanfieldenergy}) shows, on the mean field level all uniform ferromagnetic states in the O(3) order parameter manifold remain degenerate for $J_z=1$. Similarly, the mean field energy in the stripy $xy$ phases is still invariant under continuous rotations in this plane. Along the same lines, on the mean field level the order parameter in the N\'eel phase for $J_z \ge 1$ is not determined.

As we will now see   the Heisenberg-Kitaev model provides a simple example where Villain's order by disorder mechanism comes into play and restricts the order parameter to lie in a subspace of the degenerate manifold. This mechanism requires finite temperatures, where entropic contributions to the free energy become effective.  The formal procedure followed below is to integrate out the leading thermal fluctuations, and see that for certain directions of the ordered moment those fluctuations are softer and can further lower the free energy.

We shall consider explicitly the stripy region in terms of the $\tilde{\bf S}$ variables.
We introduce a slowly varying ferromagnetic order parameter field $\langle {\tilde{\bf S}}_i \rangle  \to {\bf M}({\bf r})$ of unit length
\bea
[M^x({\bf r})]^2+[M^y({\bf r})]^2+[M^z({\bf r})]^2=1,
\eea
and define gradients along the directions of the three bonds, $\nabla_{{\bf \hat{u}}_{\gamma}} = {\bf \hat{u}}_{\gamma} \cdot {\boldsymbol \nabla}$, $(\gamma=x,y,z)$ where ${\boldsymbol \nabla} = \left( \partial_x, \partial_y \right)$, with unit vectors ${\bf \hat{u}}_z ={\bf \hat{y}}$ and ${\bf \hat{u}}_{x,y} = \mp \frac{\sqrt{3}}{2} {\bf \hat{x}} - \frac{1}{2}{\bf \hat{y}}$. We set the length of these bonds to unity such that the hexagon area is $A_{\rm{hex}} =  3^{3/2}/2$ and the area of the Brilloiuin zone is $A_{BZ}=4 \pi / \sqrt{3}$.  Expanding the spin-spin interaction Eq.~(\ref{CJK}) up to second order in gradients we obtain the continuum Hamiltonian $H = \int \frac{d^2 r}{A_{\rm{hex}}} \mathcal{H}[{\bf M}] $, with
\begin{equation}
\label{cont}
\mathcal{H}[{\bf M}] = \sum_{\gamma} \frac{J_\gamma}{2} \left((1-\alpha)(\nabla_{{\bf \hat{u}}_{\gamma}} {\bf M})^2 +2(2 \alpha-1) (\nabla_{{\bf \hat{u}}_{\gamma}} M^\gamma)^2 \right).
\end{equation}
For simplicity we focus on the case $J_x=J_y=J_z=J$.

We now consider the partition function
of the continuum model Eq.~(\ref{cont}),
  \bea
Z = \int \mathcal{D} {\bf M}({\bf r})   e^{-\mathcal{H}[{\bf M}({\bf r})]/T}.
\eea
We proceed by describing the magnetization ${\bf M}({\bf r})$ in terms of fluctuations corresponding to two Goldstone modes $\pi_1({\bf r})$ and $\pi_2({\bf r})$ around a uniform magnetization ${\bf \hat{e}}$,
\bea
\label{eq:pi}
{\bf M}({\bf r}) = {\bf \hat{e}} \sqrt{1- \bar{\pi}^2({\bf r})} + \sum_{a=1,2} {\bf \hat{e}}_a \pi_a({\bf r}).
\eea
Here $\bar{\pi} = \sqrt{ \pi_1^2({\bf r})+ \pi_2^2({\bf r})}$, and the set of unit vectors $\{{\bf \hat{e}}_1,{\bf \hat{e}}_2,{\bf \hat{e}}\}$ forms an orthonormal basis. This allows to rewrite the partition function as
\begin{equation}
Z = \int \mathcal{D} {\bf \hat{e}}   \int \mathcal{D} \pi_a({\bf r}) e^{-  \mathcal{H}[{\bf \hat{e}}, \pi_a({\bf r})]/T}=\int \mathcal{D} {\bf \hat{e}}   e^{-H_{\rm{eff}}[{\bf \hat{e}}]/T},
\end{equation}
hence introducing an effective Hamiltonian of ${\bf \hat{e}}$ by integrating over the fluctuations,
\bea
e^{- H_{\rm{eff}}[{\bf \hat{e}}]/T} =  \int \mathcal{D} \pi_a({\bf r}) e^{- \mathcal{H}[{\bf \hat{e}} ,\pi_a({\bf r})]/T}.
\eea
In appendix \ref{app:m4} we compute $H_{\rm{eff}}[{\bf \hat{e}}]$ explicitly by expanding $\mathcal{H}[{\bf \hat{e}} ,\pi_a({\bf r})]$ up to quadratic order in the fluctuations $\pi_a({\bf r})$. Up to a constant and up to quadratic order in $2 \alpha-1$, we obtain the symmetry allowed anisotropic term
\bea
\label{eq:m4}
\frac{H_{\rm{eff}}}{N T} =-\frac{2}{3 } \left( 2 \alpha-1\right)^2
\left[(\hat{e}^x)^4+(\hat{e}^y)^4+(\hat{e}^z)^4\right].
\eea
This is the main result of this section. Its negative sign restricts the magnetization in the stripy phase to lie along one of the cubic axes. This term is quadratic in $\alpha-1/2$, implying the same conclusion for both sides of the point $\alpha=1/2$ in the phase diagram at $J_z=1$. Similarly, in the stripy $xy$ phases, by the same argument the magnetization is restricted to either the $x$ or $y$ cubic axes. On the classical level the N\'eel ordered phase has an equivalent description as the ferromagnet, and our order by disorder calculation implies that the N\'eel order parameter is restricted to point along one of the cubic axes.


\subsection{Numerical results}
\label{sec:numerical-results}

\newcommand{\figurebox}[1]{%
  \parbox[t][6.3cm][t]{\linewidth}{%
    \includegraphics[width=\figwidth]{{#1}}%
  }
}

\begin{figure}
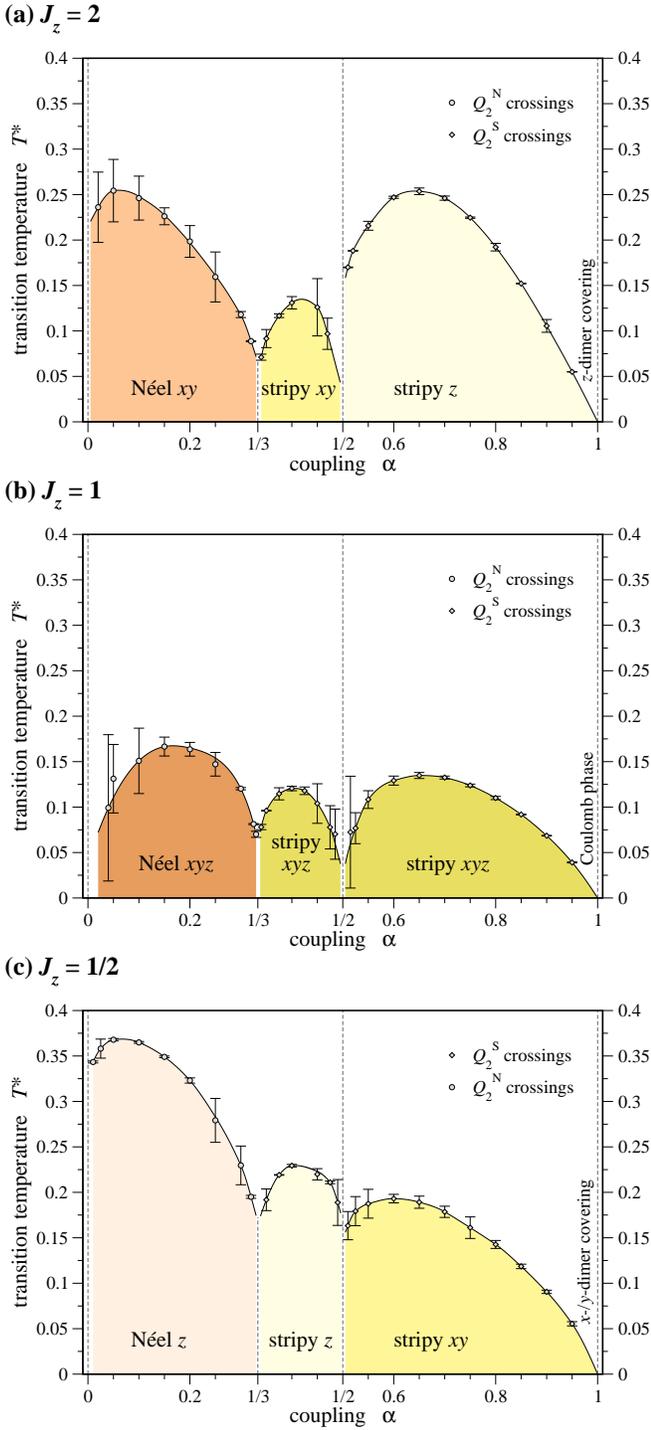

  \newlength{\figwidth}
  \setlength{\figwidth}{0.99\linewidth}
  \figurebox{{./Figures/CombinedDelta1}}
  \figurebox{{./Figures/CombinedDelta0}}
  \figurebox{{./Figures/CombinedDelta-0.5}}
  \caption{
    Finite-temperature phase diagrams for (a) $J_z>1$, (b) $J_z=1$ and
    (c) $J_z<1$.  We estimate the temperature of transition to the
    ordered phase by the intersection point of order parameter Binder
    cumulant plots $Q_2^{\text N}(T)$ for $\alpha<1/3$ and $Q_2^{\text
      S}(T)$ for $\alpha>1/3$ averaged over multiple pairs of lattice
    sizes $L$.  See Fig.~\ref{fig:binderintersect} for example data
    that went into this calculation.  The dashed lines at $\alpha=0,
    1/2$ indicate the parametrizations for which the Heisenberg-Kitaev
    model is $O(3)$ symmetric and as a consequence of the
    Mermin-Wagner theorem is not expected to display
    finite-temperature transitions in good agreement with our
    numerical analysis.  The degenerate dimer-covering states at
    $\alpha=1$ also do not undergo any phase transition for $T>0$.}
  \label{fig:transitiontemperatures}
\end{figure}

\begin{figure}
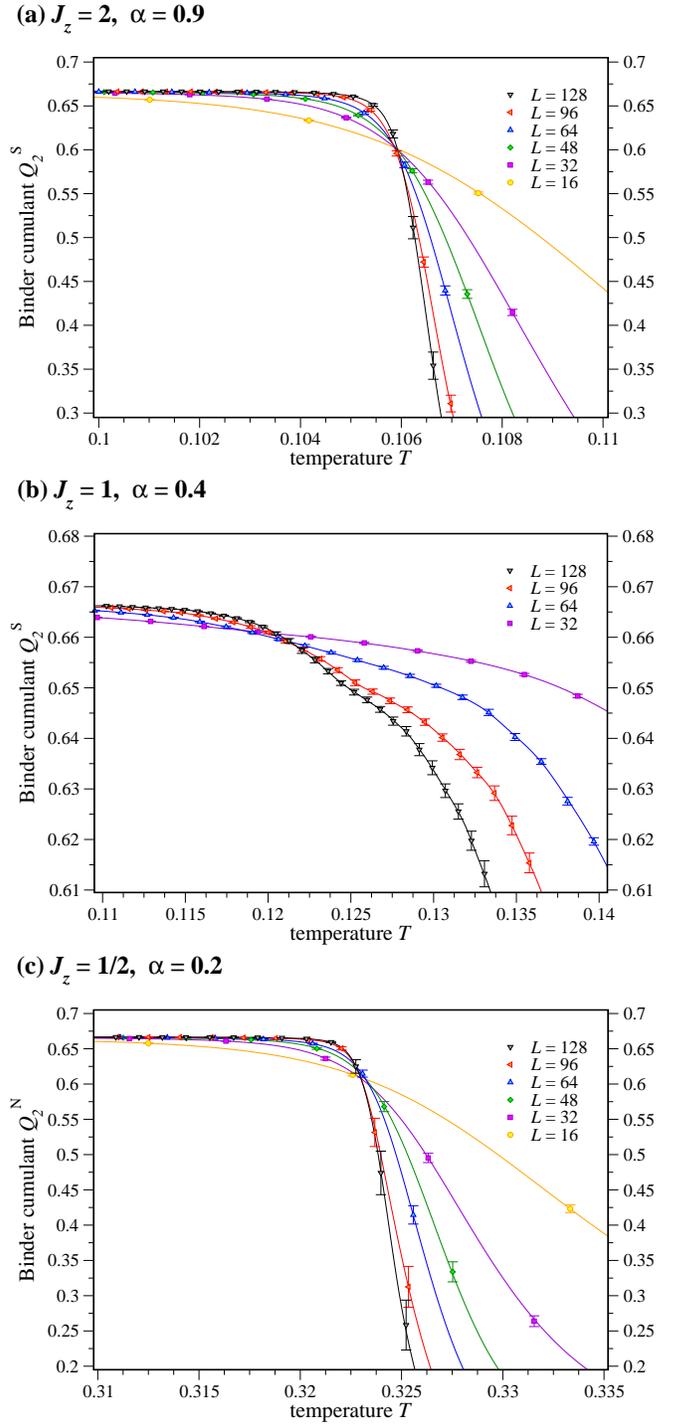

  \setlength{\figwidth}{0.99\linewidth}
  \begin{center}
    \figurebox{{./Figures/CrossingDelta1}}
    \figurebox{{./Figures/CrossingDelta0}}
    \figurebox{{./Figures/CrossingDelta-0.5}}
  \end{center}
  \caption{Binder cumulant curves of the order parameter evaluated     over $T$ for various system sizes $L$ close to their crossing,     which gives an estimate of the transition temperature for several     example parameter sets.  Shown are the cumulant of $\vmS$ for the     transitions to (a) the stripy $z$ phase and (b) the stripy $xyz$     phase as well as (c) the cumulant of $\vmN$ for the transition to     the N\'eel $z$ phase.  Symbols with error bars are single     temperature data, while continuous lines are interpolated by     multiple histogram reweighting. }
  \label{fig:binderintersect}
\end{figure}

\begin{figure}
  \includegraphics[width=.7\columnwidth]{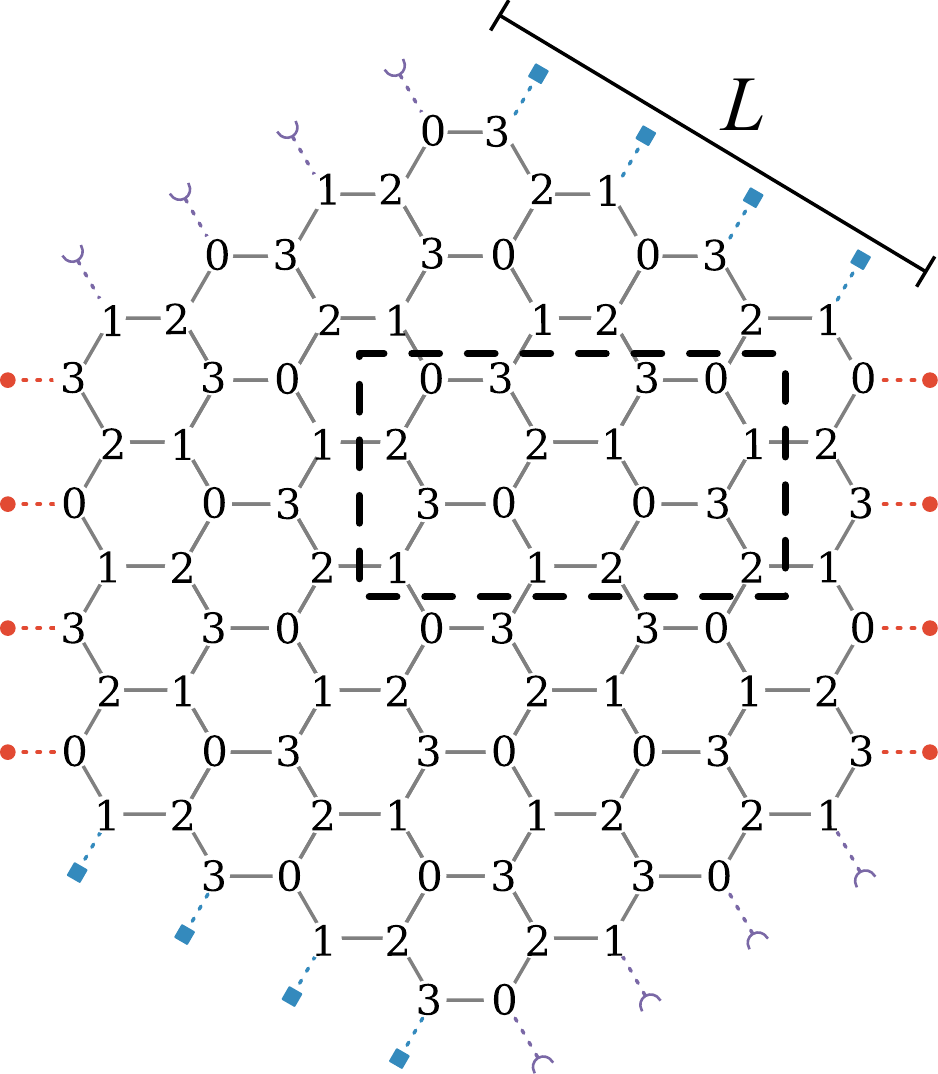}
  \caption{$L=4$ example of the finite lattices used in our Monte
    Carlo simulations, where opposing boundaries are periodic as
    indicated.  The numbers show the division into the sublattices
    used in the definition of the order parameter $\vmS$ in
    Eq.~\eqref{eq:mc1}, which matches the supercell in
    Fig.~\ref{fg:supercell}.}
  \label{fig:mclattice}
\end{figure}

Our analysis of the classical Heisenberg-Kitaev model is complemented
by an extensive finite-temperature Monte Carlo study.  In our
simulations the classical spins ${\bf S}_i$ are situated on the
vertices of hexagon-shaped clusters with periodic boundary conditions,
which realize the $C_3$ symmetry of the honeycomb lattice and allow to
observe unbiasedly all possible orientations in the stripy phases; see
Fig.~\ref{fig:mclattice}.  A cluster with a side length of $L$
plaquettes contains $N=6L^2$ sites.

We apply the standard Metropolis algorithm\cite{Newman1999,Janke2008}
with two different types of proposed moves: In one lattice sweep we
first perform local updates of each individual spin, where the new
orientation is chosen from an angular region around the old
orientation, which has been tuned in such a way during thermalization
that acceptance ratios of $50\%$ are maintained at all temperatures.
In a second stage we then propose $3N$ ``bond-flip'' moves.  In one of
these moves we choose a random pair of nearest-neighbor sites together
with their associated bond-direction $\langle i,j \rangle^{\gamma}$.
Then for the spins at both sites we reverse the sign of the
spin-component linked via that bond in the Kitaev interaction:
$S_i^{\gamma} \to -S_i^{\gamma}$ and $S_j^{\gamma} \to -S_j^{\gamma}$,
whereas the other components are not modified.  While the bond-flip
update would not be ergodic on its own, in combination with the
single-spin update it greatly accelerates simulation dynamics in the
stripy phases, vastly facilitating equilibration.

To further improve ergodicity we combine these canonical updates with
a parallel-tempering scheme.\cite{Geyer1991,HukushimaNemoto}  Here we
simulate multiple replicas of the spin system concurrently at
different temperatures and exchange configurations between them in a
controlled manner that satisfies detailed balance.  In this way short
autocorrelation times at high temperatures can be exploited to easily
overcome free energy barriers at low temperatures, and we can reach
all relevant regions of phase space in a single simulation regardless
of initial conditions.

We measure two vector order parameters to distinguish between
different antiferromagnetic spin-alignments:
\begin{align}
  \label{eq:mc1}
  \vmN &= \frac{1}{N}\left( \sum_{i \in a} {\bf S}_i - \sum_{i \in
      b} {\bf S}_i \right) \quad\text{and} \notag \\
  \vmS &= \frac{1}{2N}\left( \sum_{i \in 0}{\bf S}_i + \sum_{i \in 1}
    {\bf S}_i - \sum_{i \in 2}{\bf S}_i + \sum_{i \in 3}{\bf S}_i
  \right) .
\end{align}
Here $a$ and $b$ stand for the two sublattices of the honeycomb
lattice, while the four honeycomb sublattices formed by the sites of
the different types of the supercell of Fig.~\ref{fg:supercell} are
denoted by $0$, $1$, $2$ and $3$.  Fig.~\ref{fig:mclattice} shows how
these sublattices are assigned in our finite lattices.  $|\vmN| = 1$
corresponds to perfect N\'eel order, while $|\vmS| = 1$ is realized
for perfect stripy order.  The preferred orientations of the
magnetization vectors $\vmN$ and $\vmS$ reflect which ordering
directions are possible in the different N\'eel and stripy phases.  In
Eq.~\eqref{eq:mc1} we have chosen an asymmetric definition of the
order parameter $\vmS$, where one of the sublattice magnetizations is
counted negative and three are counted positive.  With this definition
$\vmS$ is simultaneously an order parameter for the stripy $x$, $y$
and $z$ phases on the same lattice.\footnote{Note that
  Refs.~\onlinecite{perkins12,perkins13} use an alternative definition
  of the stripy-order parameter, which is specified on a choice of
  sublattices different from ours.} By measuring histograms of the
components of $\vmN$ and $\vmS$ we were able to verify the analytical
arguments of section~\ref{se:orderbydisorder}.  We obtain planar
representations of $\vmN$ and $\vmS$ by mapping the three Cartesian
basis vectors to the complex plane as in ${\bf \hat{e}}_x \to \exp(7 i
\pi /6)$, ${\bf \hat{e}}_y \to \exp(11 i \pi /6)$ and ${\bf \hat{e}}_z
\to \exp(i \pi /2)$ and show the resulting histograms as insets in the
phase diagram of Fig.~\ref{fig:distortedphasediagram}.  Both the
carefully chosen shape of the finite lattices and the
parallel-tempering algorithm are essential tools allowing us to fully
explore configuration space in our simulations as reflected in these
histograms.

Recently Price and Perkins~\cite{perkins12,perkins13} studied the
undistorted, $C_3$-symmetric classical Heisenberg-Kitaev model at
finite temperature. Following their analysis we study the Binder
cumulants of the absolute valued order parameters
\begin{align}
  \label{eq:mc2}
  \QN = 1 - \frac{1}{3}\frac{\left\langle \mN^4 \right\rangle}{
    \left\langle \mN^2 \right\rangle^2} \quad\text{and}\quad
  \QS = 1 - \frac{1}{3}\frac{\left\langle \mS^4 \right\rangle}{
    \left\langle \mS^2 \right\rangle^2}
\end{align}
in order to pinpoint the precise temperature of the transitions into the
ordered phases. At criticality their values depend only weakly on the
system size.  Hence the intersection point of $\QN(T)$ or $\QS(T)$
curves evaluated for different $L$ gives a good estimate of the
critical temperature.

Interestingly, Price and Perkins found that for $\alpha \ne 0,1/2,1$,
the entrance to the ordered phases (N\'eel or stripy) from the high-temperature
paramagnetic phase undergoes {\em two} consecutive phase
transitions, via a small sliver of a critical Kosterlitz-Thouless
phase.  In this intermediate phase, the effective model is a 6-state
clock model, corresponding to the 6 possible stripy or N\'eel phases,
where an effective U(1) symmetry emerges.  However, for the distorted
model there are only 2 or 4 degenerate stripy or N\'eel phases.  In
this case the intermediate U(1) symmetric phase is not expected
\cite{kadanoff} and no evidence of it is found in our numerical
analysis.

We apply standard multiple histogram
reweighting techniques\cite{Ferrenberg1989,Chodera2007} to the
temperature-sorted observable time series in combination with
numerical minimization routines\cite{NumericalRecipees} to find the
intersection points for systems of different sizes up to $L=128$.
Statistical uncertainties are estimated by performing the entire
analysis on jackknife resampled data sets.\cite{Efron1982}  Plots of
Binder cumulants close to their crossing points are given in
Fig.~\ref{fig:binderintersect} for several parameter sets.  We average
over the results for different values of $L$ to estimate the
transition temperatures $T^{*}$ shown in
Fig.~\ref{fig:transitiontemperatures}.

\begin{figure*}
  \centering
  \includegraphics[width=0.95\linewidth]{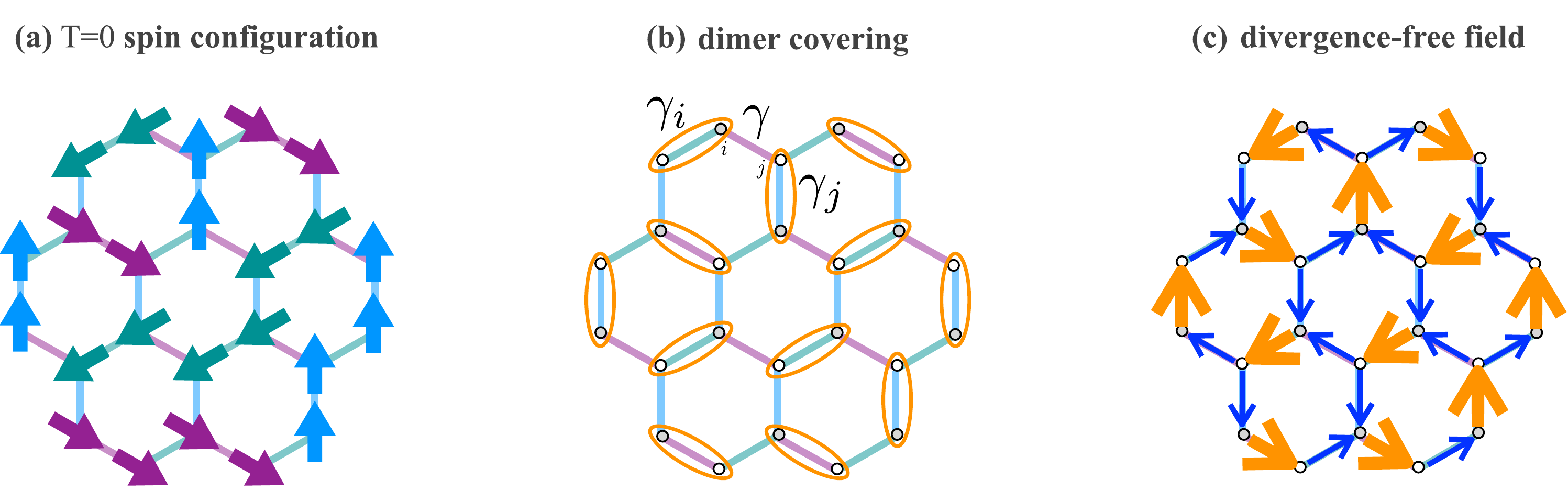}
  \caption{\label{fg:dimcover} A zero-temperature configuration of the
    $C_3$ symmetric Kitaev model is a generic dimer covering state, to
    which we can associate a divergence-free field.}
\end{figure*}

For the symmetric case $J_z=1$ our approach resolves the lower of the two
transition temperatures of the analysis of Ref.~\onlinecite{perkins13}.
For the distorted model, only  a single transition is expected
as argued above.  We associate this transition with the crossing
point of the Binder cumulant curves of different sizes as shown in the
examples in Fig.~\ref{fig:binderintersect}.


\subsection{Emergent magnetostatics in the Kitaev limit}
\label{se:Coulomb}


Before concluding our discussion of the classical Heisenberg-Kitaev model, we will briefly discuss the physics of the Kitaev limit $(\alpha=1)$. While its quantum mechanical counterpart is well known as a paradigmatic, exactly solvable spin model harboring various spin liquid ground states, the classical Kitaev model certainly deserves some attention as well. In its undistorted form $(J_x=J_y=J_z=1)$ it is one of the simplest, analytically tractable classical spin models that evades a thermal phase transition and harbors a classical spin liquid state, which at zero temperature exhibits an extensive degeneracy and pair correlations decaying with a characteristic power-law \cite{chandra}. These zero-temperature features can be traced back to an effective description in terms of emergent magnetostatics -- an example of a so-called Coulomb gas \cite{henley}.
We will briefly review the arguments showing the origin of this emergent spin liquid in the classical Kitaev model in the following with a more detailed and self-consistent account being given in appendix  \ref{app:Coulomb}. We then discuss the effect of finite distortions, which lead to a (partial) lifting of the zero-temperature degeneracy and a break-down of the Coulomb correlations. However, characteristic remnants of the Coulomb description remain as signatures in the low-temperature specific heat as we detail in the subsequent subsection.

As noted earlier the undistorted classical Kitaev model incorporates a high level of exchange frustration with each spin being subject to competing magnetic exchanges that equally favor alignment along one of the three orthogonal axes of a classical $O(3)$ Heisenberg spin.
As one approaches the zero-temperature limit of this model it is easy to see \cite{chandra} that the total energy of the system can be minimized by spin configurations where spins align in a pairwise fashion along one of the three easy axes of the magnetic exchange, i.e. the one favored by the bond between the two spins forming a pair. An example of such a spin configuration is illustrated in Fig.~\ref{fg:dimcover}(a). Since every spin is part of precisely one such aligned pair, we can identify each pair of aligned spins with a `dimer'. As a consequence, any such energy minimizing spin configuration can be mapped to a hardcore {\em dimer covering} of the honeycomb lattice as illustrated in Fig.~\ref{fg:dimcover}(b) where every site (spin) is part of precisely one dimer. This mapping allows two immediate conclusions. First, it is well known since the early work of Wannier~\cite{wannier}, Kasteleyn~\cite{kasteleyn}, and Elser~\cite{elser} that the number of dimer coverings on the hexagonal lattice grows exponentially in the system size and as thus we can immediately estimate the zero-temperature degeneracy of the spin model. Second, it has long been appreciated~\cite{henley} that the hard-core dimer constraint on a bipartite lattice allows a mapping of any dimer covering to a {\em divergence-free field} configuration, which is schematically illustrated in Fig.~\ref{fg:dimcover}(c).
It is precisely this description of the zero-temperature spin configurations in terms of a divergence-free magnetic field that allows to draw the connection to an emergent {\em Coulomb gas} description.
The latter is well known to give rise to power-law correlations, which translated back to the original spin model are pair correlations of the form
\[
    \left\langle \left( S_i^z \right)^2 \cdot \left( S_{i+r}^z \right)^2 \right\rangle \propto \frac{1}{r^2} \,.
\]
For a detailed and self-consistent description of the Coulomb gas formulation of the zero-temperature classical Kitaev model we refer the reader to appendix \ref{app:Coulomb}.

When introducing distortions of the exchange couplings the extensive degeneracy of zero-temperature states is immediately lifted. For $J_z>1$ two spin configurations are singled out where spins align along the $z$-direction again in a pairwise fashion -- with both states being mapped to an identical dimer covering as illustrated on the left-hand side in Fig.~\ref{fg:dimerz}. As a consequence, the spin liquid physics disappears entirely and the system undergoes a conventional $Z_2$ symmetry breaking thermal phase transition into one of the two states.
For $J_z<1$ a different picture emerges. While the extensive zero-temperature degeneracy is still lifted, the system retains a subextensive degeneracy down to zero temperature where the spins align in pair-wise fashion along the zig-zag chains spanned by the $x$ and $y$-bonds as illustrated on the right-hand side in Fig.~\ref{fg:dimerz}. The consequence of this lifting again is the loss of Coulomb correlations, but the system still evades a conventional ordering transition down to zero temperature with characteristic features arising for instance in the specific heat as discussed in the next subsection.


\subsubsection*{Specific heat and zero modes}
\label{se:c}
One characteristic feature of the extensive manifold of zero-temperature spin configurations is that it gives rise to certain soft fluctuations called {\em zero modes}. Following the pioneering work of Chalker \emph{et al.}~\cite{chalker}, we show in the remainder of this section that these zero modes reduce the  specific heat in its $T \to 0$ limit in a universal way -- a characteristic signature that as we show can easily be tracked by numerical simulations of the classical spin model.

To start our discussion of the analytical arguments we consider fluctuations around a given dimer covering or spin configuration, respectively.  Each spin $i$ belonging to a dimer on a $\gamma$-bond gives rise to possible fluctuations in the two directions orthogonal to $\gamma$. For example for a spin belonging to a $z$ dimer and pointing along $+z$ we write
\bea
{\bf S}_i = (\epsilon_{i}^x,\epsilon_{i}^y,\sqrt{1-{\epsilon_{i}^x}^2-{\epsilon_{i}^y}^2}).
\eea
The fluctuations in the $x$ and $y$ directions influence also the $z$ component due to the unit constraint $|{\bf S}_i|=1$.

\begin{figure}[t]
\begin{center}
  \includegraphics*[width=35mm]{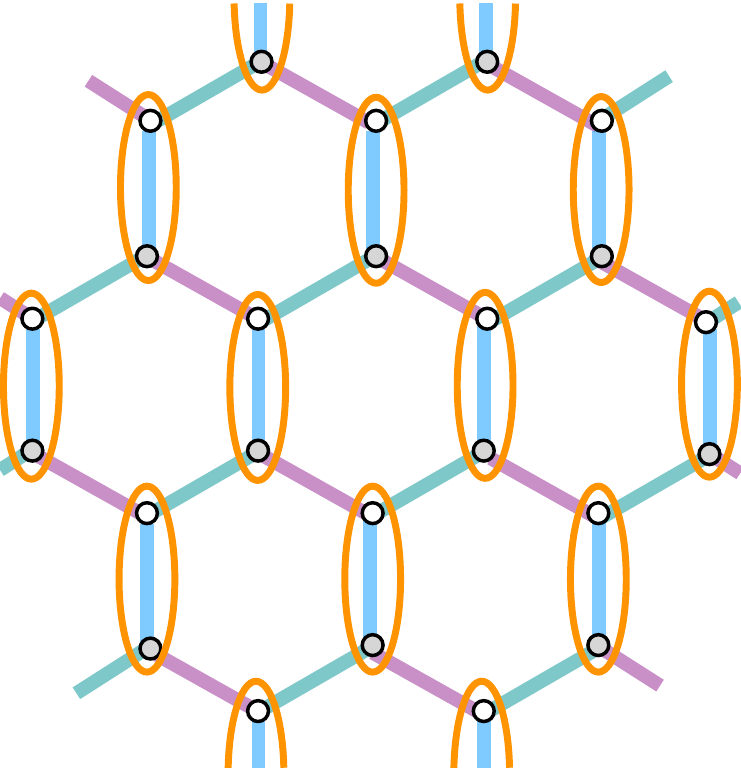}
  \hspace{10mm}
  \includegraphics*[width=35mm]{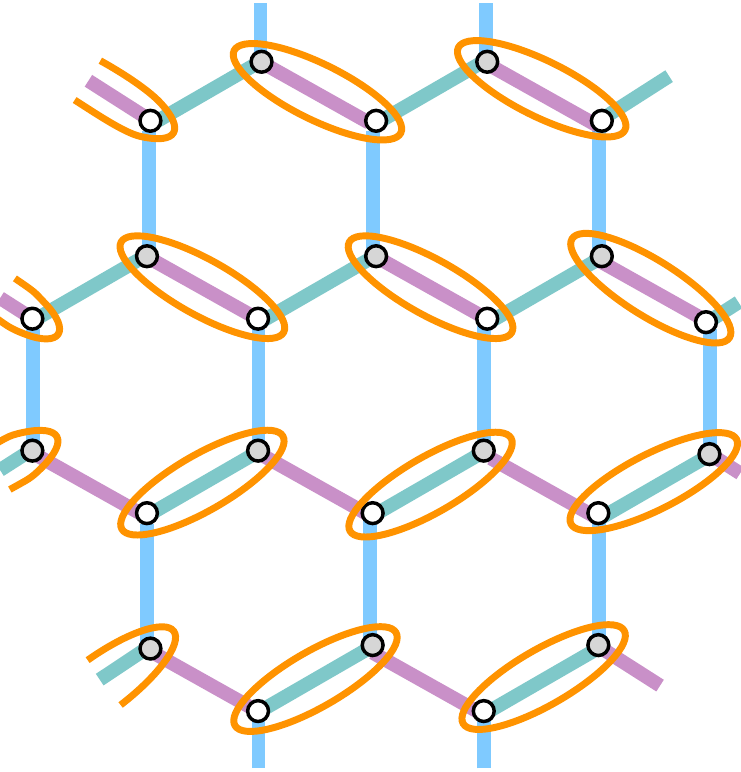}
\caption{Left: The preferable dimer covering state for dominating $J_z$.
  Right: Typical dimer covering states for $J_x = J_y>J_z$
  \label{fg:dimerz}
}
\end{center}
\end{figure}

Let $\mathcal{D}$ denote the set of dimerized bonds. For $\langle i,j \rangle \in \mathcal{D}$, and assuming for simplicity that this is a $z$ type bond, the Kitaev spin-spin interaction reads
\bea
-J_z S_i^z S_j^z|_{\langle i,j \rangle \in \mathcal{D}} &=& -J_z\sqrt{1-{\epsilon_{i}^x}^2-{\epsilon_{i}^y}^2}\sqrt{1-{\epsilon_{j}^x}^2-{\epsilon_{j}^y}^2} \nonumber \\
 &=& -J_z + \frac{J_z}{2} ({\epsilon_{i}^x}^2+{\epsilon_{i}^y}^2+{\epsilon_{j}^x}^2+{\epsilon_{j}^y}^2)\nonumber \\&+&\mathcal{O}(\epsilon^4).
\eea
We see that up to quadratic order, fluctuations do not interact across
dimerized bonds (no $\epsilon_i \epsilon_j$ coupling terms for
$\langle i,j \rangle \in \mathcal{D}$). On the other hand, for a non-dimerized bond $\gamma$ [see Fig.~\ref{fg:dimcover}(b)] the Kitaev interaction reads
\bea
-J_\gamma S^\gamma_i S^\gamma_j|_{\langle i,j \rangle \notin \mathcal{D}} &=& -J_\gamma \epsilon_i^\gamma \epsilon_j^\gamma.
\eea
Thus, expanding the Hamiltonian in $\epsilon$ to quadratic order, the fluctuation corrections consist of decoupled terms which live on the non-dimerized bonds and read
\bea
\label{Hnondimer}
H^{(2)} = \sum_{\langle i,j \rangle_{\gamma} \notin \mathcal{D}} h(\epsilon_i^\gamma,\epsilon_j^\gamma),
\eea
where
\bea
h(\epsilon_i^\gamma,\epsilon_j^\gamma) =  - J_\gamma \epsilon_i^\gamma \epsilon_j^\gamma +\frac{1}{2}(J_{\gamma_i} {\epsilon_i^\gamma}^2+J_{\gamma_j} {\epsilon_j^\gamma}^2).
\eea
\begin{figure}
  \includegraphics[width=\linewidth]{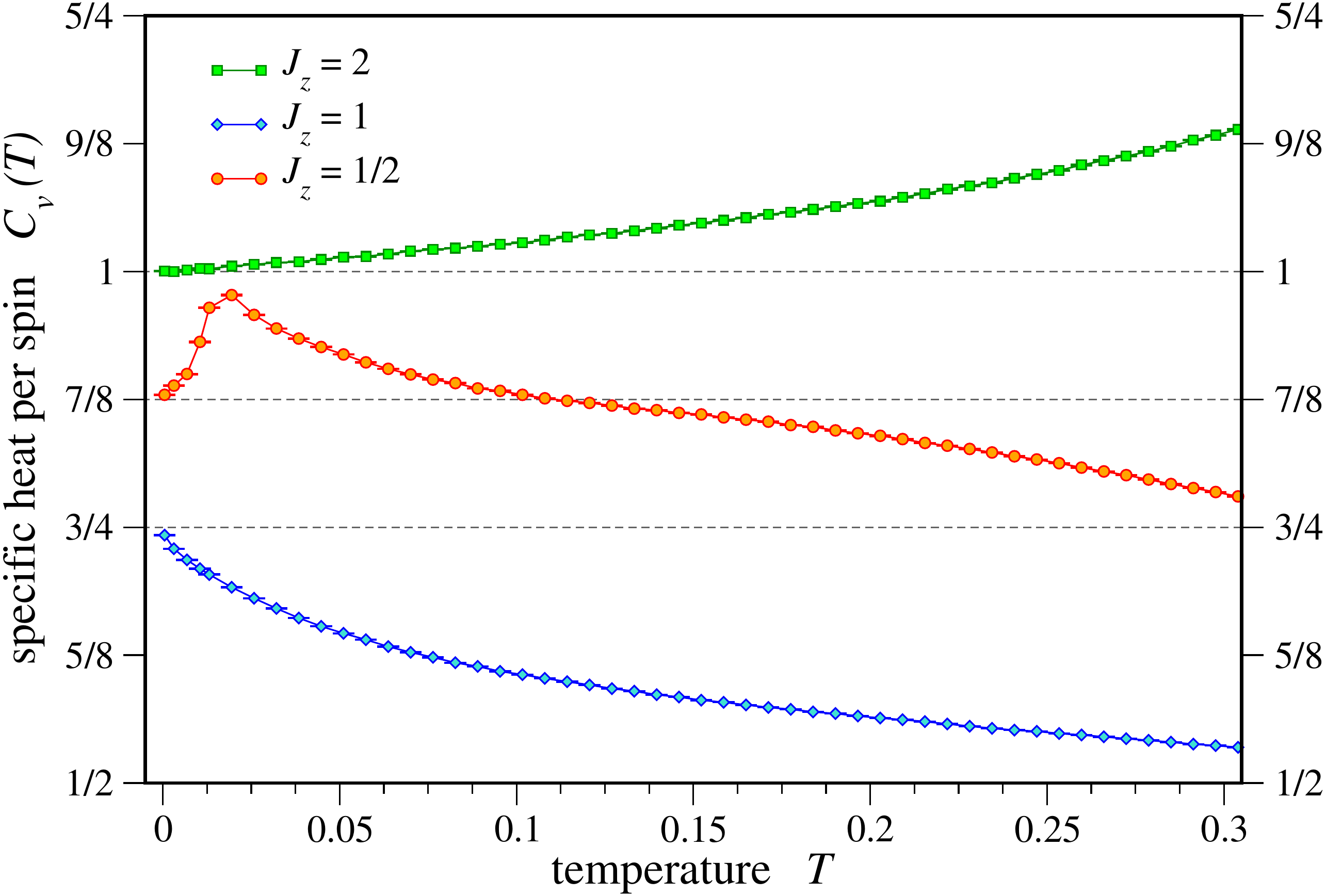}
  \caption{Low temperature behavior of the specific heat per spin
    $C_v(T)$ in the classical Kitaev model with different distortions.
    Shown are Monte Carlo results obtained at temperatures $T \ge
    1/2000$ demonstrating that in the limit of $T \to 0$ one finds
    $C_v^{J_z=1} \to 3/4$, $C_v^{J_z>1} \to 1$ and $C_v^{J_z<1} \to
    7/8$.  The data has been obtained for systems of side length
    $L=16$.}
  \label{fig:cvtozero}
\end{figure}
Interestingly, for $J_x = J_y = J_z$,
\bea
 \label{h0}
 h(\epsilon_i^\gamma,\epsilon_j^\gamma) =  -\frac{J_z}{2} (\epsilon_i^\gamma-\epsilon_j^\gamma)^2.
 \eea
This implies the existence of a zero mode: $(\epsilon_i^\gamma+\epsilon_j^\gamma)$ does not appear in $H^{(2)}$. This zero mode has been identified~\cite{shankar} to be a sliding degree of freedom of the dimer covering states. For low enough temperatures fluctuations become small and the partition function becomes
\bea
Z \cong \int \mathcal{D}(\{ \epsilon \})e^{-\frac{H^{(2)}(\{ \epsilon \})}{T}}.
\eea
For any quadratic eigenmode $\epsilon$, with energy $E =c_2 \epsilon^2$, the contribution to the specific heat then becomes
\bea
\label{eq:eq}
C_v = \frac{d}{dT} \frac{\int d \epsilon (c_2 \epsilon^2) e^{-\frac{ c_2 \epsilon^2}{T}}}{\int d \epsilon e^{-\frac{ c_2 \epsilon^2}{T}}}
=\frac{\int dx x^2 e^{-x^2}}{\int dx  e^{-x^2}}= \frac{1}{2},
\eea
independent of the coefficient $c_2$. However, in our system we have to further consider the contributions of the zero modes. For those modes we need to go to quartic order, i.e. $E =c_4 \epsilon^4$, for which the contribution to the specific heat can be estimated to be
\bea
C_v = \frac{d}{dT} \frac{\int d \epsilon (c_4 \epsilon^4) e^{-\frac{ c_4 \epsilon^4}{T}}}{\int d \epsilon e^{-\frac{ c_4 \epsilon^4}{T}}}
=\frac{\int dx x^4 e^{-x^4}}{\int dx  e^{-x^4}}= \frac{1}{4},
\eea
again independent of the coefficient $c_4$. In a standard state without zero modes (such as a ferromagnetic state) we would have two quadratic modes ($\epsilon^x_i$ and $\epsilon^y_i$) per spin. This would give the zero temperature value of the specific heat per spin
\bea
C_v^{\rm{ferro}}(T \to 0) =\frac{1}{2}+ \frac{1}{2} = 1.
\eea
However, in the Coulomb phase of the classical Kitaev model, we have only one zero mode for each quadratic mode, hence
\bea
C_v^{J_z=1}(T \to 0) =\frac{1}{2}+ \frac{1}{4} = \frac{3}{4}.
\eea

We now consider the effect of a finite distortion, i.e. $J_z \ne J_x = J_y$, which splits the degeneracy of the various dimer covering states.  For $J_z>J_x = J_y$, namely $J_z>1$, the dimer covering with only $z$-dimers has the lowest energy; see Fig.~\ref{fg:dimerz}.

At the same time, fluctuations around this state are described by Eq.~(\ref{Hnondimer}), which can be written as
\bea
\label{h1}
h(\epsilon_i^\gamma,\epsilon_j^\gamma) = \frac{3(J_z-1)}{8}(\epsilon_i^\gamma+\epsilon_j^\gamma)^2+ \frac{J_z+3}{8}(\epsilon_i^\gamma-\epsilon_j^\gamma)^2.
\eea
For $J_z>1$ the two coefficients in this equation are positive, leaving no zero modes. Hence
\bea
C_v^{J_z>1}(T \to 0) =\frac{1}{2}+ \frac{1}{2} = 1.
\eea

For $J_z<J_x = J_y$ the dimers cover $x$ or $y$ bonds in the ground state; see Fig.~\ref{fg:dimerz}. Now consider fluctuations around these 1D covering states. The Hamiltonian for the fluctuations is the same as Eq.~(\ref{Hnondimer}), but now there are two types of non-dimerized bonds. For $\langle i,j \rangle_{\gamma} \notin \mathcal{D}$ with $\gamma=x$ or $y$, $h$ has the form of Eq.~(\ref{h0}), implying a zero mode. But for $\langle i,j \rangle_{\gamma} \notin \mathcal{D}$ with $\gamma=z$, the Hamiltonian $h$ has the form of Eq.~(\ref{h1}), implying no zero mode. As a result the specific heat per spin becomes
\bea
C_v^{J_z<1}(T \to 0) =\frac{1}{2}\left(\frac{1}{2}+\frac{1}{4} \right)+\frac{1}{2}\left(\frac{1}{2}+\frac{1}{2} \right) = \frac{7}{8}.
\eea
Our Monte Carlo calculations, summarized in Fig.~\ref{fig:cvtozero},
nicely reproduce these fractions and are thus able to pinpoint the different
constraints on the dimer covering states underlying the Coulomb gas.

%
%

\section{Quantum Heisenberg-Kitaev model}
\label{sec:quantum}

\begin{figure*}[t]
  \includegraphics[width=.8\linewidth]{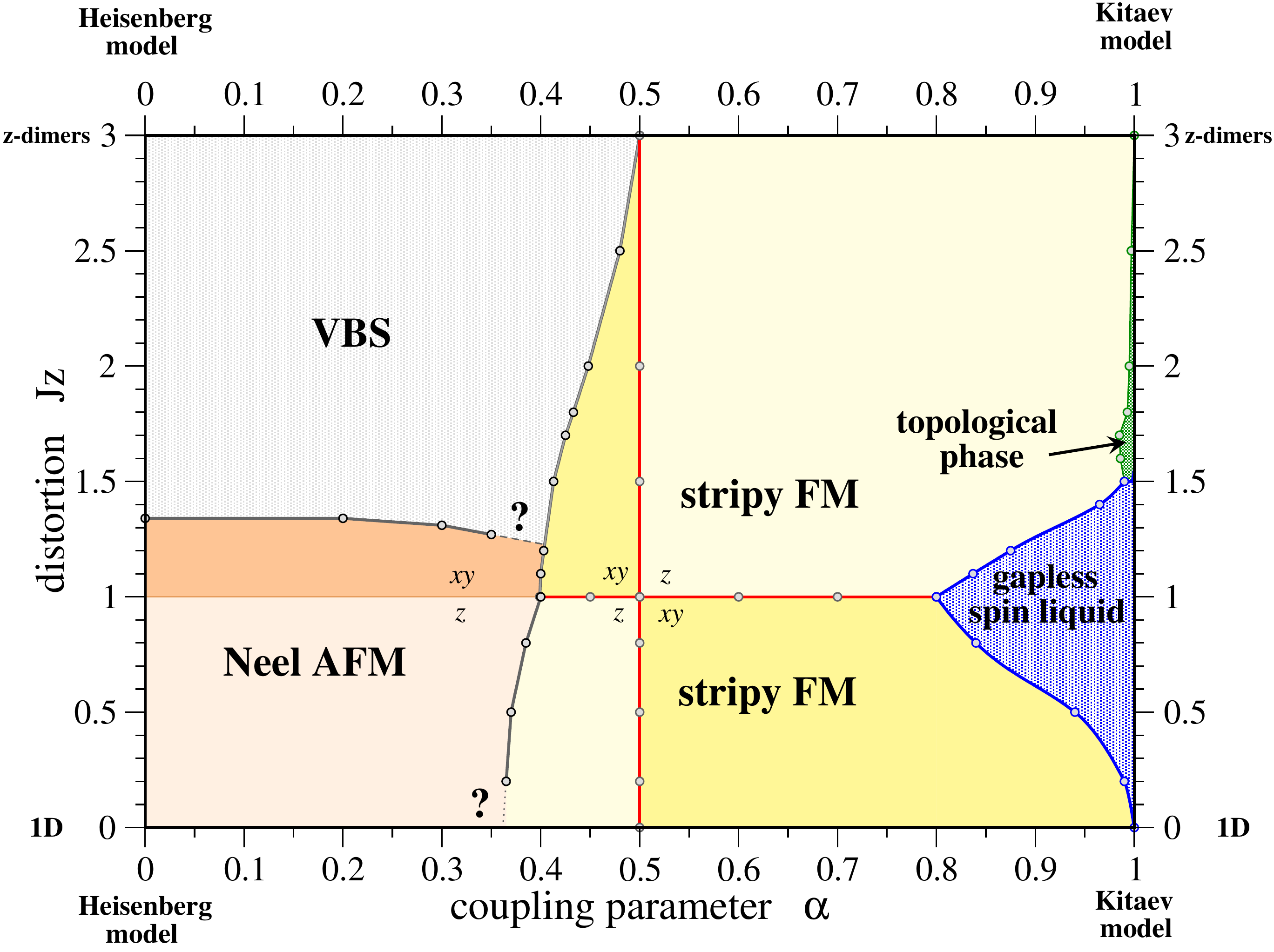}
\caption{Phase diagram of the distorted quantum Heisenberg-Kitaev model.}
  \label{fig:QuantumPhaseDiagram}
\end{figure*}

We now turn to a discussion of the quantum version of the distorted Heisenberg-Kitaev model, i.e. we again consider the Hamiltonian
\begin{eqnarray}
   H & = &  (1-\alpha) \mathcal{H}_{\rm Heisenberg} - 2\alpha \mathcal{H}_{\rm Kitaev} \nonumber \\
      & = &  \sum_{\langle ij \rangle, \gamma} J_\gamma\, \left( (1-\alpha)\, {\bf S}_i {\bf S}_{j}  - 2\alpha\,  S^\gamma_i S^\gamma_{j} \right) \,,
\end{eqnarray}
where the spins ${\bf S}_i$ are now quantum mechanical $SU(2)$ spin-1/2 degrees of freedom. (In our convention ${\bf S}_i$ are represented by Pauli matrices $(S_i^\gamma)^2=1$.) The exchange parameter $0 < \alpha < 1$ again interpolates between the antiferromagnetic Heisenberg model $(\alpha=0)$ and the ferromagnetic Kitaev model $(\alpha=1)$ and the distortion of the exchange couplings is parametrized by $0<J_z<3$ with the simultaneous conditions that all three spin exchange couplings add up to a constant, i.e. $J_x+J_y+J_z=3$, and $J_x=J_y$. The case $J_z=1$ then corresponds to the undistorted situation where the spin exchange along all three bonds has equal magnitude, i.e. $J_x=J_y=J_z$. The limit $J_z=3$ $(J_x=J_y=0)$ corresponds to decoupled dimers on the $z$-bonds, while the opposite limit of $J_z=0$ $(J_x=J_y=3/2)$ corresponds to decoupled zig-zag chains along the $x$- and $y$-bonds.

When exploring the $(\alpha, J_z)$-parameter space we find that the above model not only harbors quantum analogues of all classically ordered states, but exhibits a number of additional genuinely quantum states including a valence-bond solid and two spin-orbital liquid phases, which both extend well beyond the well-studied Kitaev limit of the quantum model. In fact, one of the more interesting features of the extended phase diagram of the quantum Heisenberg-Kitaev model is the possible occurrence of unconventional continuous phase transitions between these gapped and gapless spin-orbital liquid phases and conventionally ordered states.

In the following, we will first discuss the general quantum phase diagram of the distorted Heisenberg-Kitaev model and the numerical simulations underlying its determination and then focus our discussion on the possibly interesting quantum critical behavior associated with the phase transition out of one of the spin-orbital liquid phases.

\subsection{Phase diagram of the quantum model}

The phase diagram of the quantum Heisenberg-Kitaev model in the presence of exchange distortions is summarized in Fig.~\ref{fig:QuantumPhaseDiagram}. Similar to the classical model we find an extended N\'eel ordered phase around the Heisenberg limit which upon distorting the exchange interactions undergoes a quantum order-by-disorder transition locking the spin orientation in the ordered phases to the $z$ ($x$ or $y$) direction for $J_z<1$ ($J_z>1$), respectively. For $J_z \gtrsim 1.35$ the system undergoes a transition into a valence bond solid (VBS), which adiabatically connects to the limit of isolated dimer singlets on the $z$-bonds in the limit $J_z=3$ (and $\alpha<1/2$).

For $\alpha=1/2$ the quantum model exhibits an $SU(2)$ symmetry that is again rooted in the observation that for this ratio of the Heisenberg and Kitaev couplings the model can be mapped via the four-sublattice basis transformation illustrated in Fig.~\ref{fg:supercell} to a ferromagnetic Heisenberg model. In fact, such a mapping exists for all values of the distortion $J_z$, i.e. the quantum model exhibits an entire $SU(2)$ symmetric line for $\alpha=1/2$. In the four-sublattice rotated basis the ground state of the quantum model is a simple ferromagnet for $\alpha=1/2$, which transformed back into the original basis becomes a `stripy ferromagnet' akin to the illustrations in Fig.~\ref{fg:stripy}. In the undistorted case $(J_z=1)$ the ground state is sixfold degenerate with the six possible stripy states of Fig.~\ref{fg:stripy} having equal weight in the ground state. This picture changes immediately upon moving away from the $\alpha=1/2$ line and distorting the exchange couplings. Again a quantum order-by-disorder transition (detailed in appendix \ref{app:QuantumOBDO}) selects a subset of these six stripy states with four different phases emerging around the undistorted $(\alpha=1/2, J_z=1)$ point in the middle of our phase diagram in Fig.~\ref{fig:QuantumPhaseDiagram}.
In complete analogy to the classical model, a subset of two stripy FM states locking the spins into the $z$-direction is selected for $(\alpha>1/2,J_z>1)$ as well as for $(\alpha<1/2,J_z<1)$. For the other two quadrants $(\alpha<1/2,J_z>1)$ and $(\alpha>1/2,J_z<1)$ the opposite subset of four stripy FM states with the spins locking into either the $x$ or $y$-directions are selected by the quantum order-by-disorder mechanism, see appendix \ref{app:QuantumOBDO} for details.

Arguably the most interesting phases in our phase diagram are the two spin liquid phases emerging for dominating Kitaev couplings. For the undistorted Heisenberg-Kitaev model it was previously established \cite{chaloupka,jiang} that the stripy FM phase gives way to a gapless spin liquid phase for $\alpha \approx 0.8$, i.e. Kitaev couplings which are about 8 times larger than the isotropic Heisenberg exchange. This gapless spin liquid phase remains stable when introducing an exchange distortion $J_z \neq 1$ and is found to occupy a rather extended regime in the $(\alpha, J_z)$-parameter space as illustrated in Fig.~\ref{fig:QuantumPhaseDiagram}.
For the pure Kitaev model it is well known \cite{Kitaev} that the gapless spin liquid can be gapped out into a topological spin liquid if one introduces an exchange distortion that renders one of the three coupling exchanges dominant, i.e. $J_z \geq 3/2$ in our notation, see Fig.~\ref{fig:kitaev}(c). Upon including a Heisenberg exchange this gapped phase must remain stable for a finite parameter regime -- however, since the gap itself is rather small the regime occupied by this topological spin liquid in our $(\alpha, J_z)$-parameter space reduces to a small sliver as illustrated in Fig.~\ref{fig:QuantumPhaseDiagram}. We come back to a more detailed discussion of the emergence of this topological phase as well as the nature of the quantum phase transition out of this phase into the stripy phase in the next subsection.

\begin{figure}
  \includegraphics[width=\linewidth]{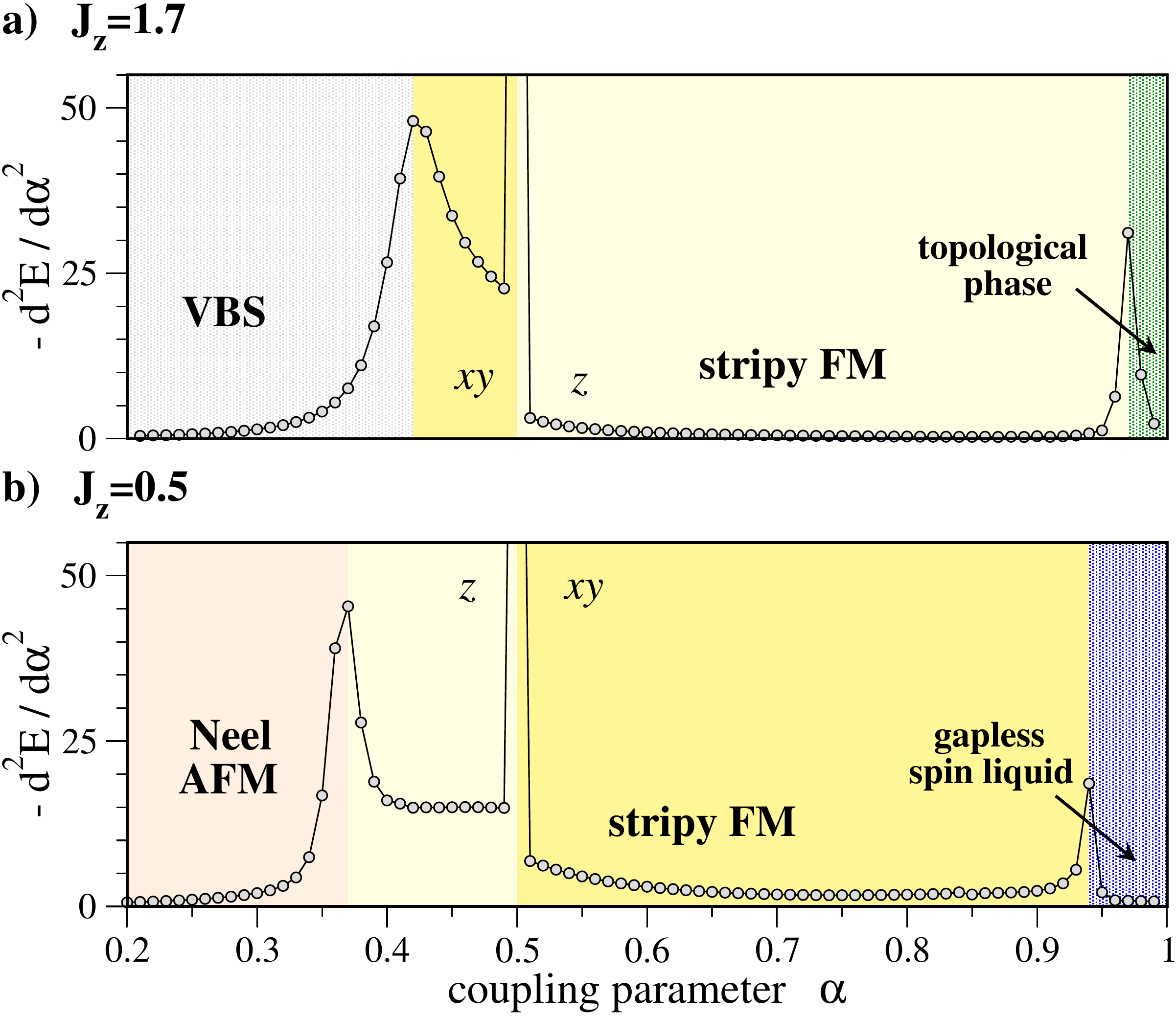}
  \caption{(Color online) Second derivative of the ground state energy density as a function of $\alpha$ for two different values of the distortion $J_z$.}
  \label{fig:energderivativeyscan}
\end{figure}

\begin{figure}
  \includegraphics[width=\linewidth]{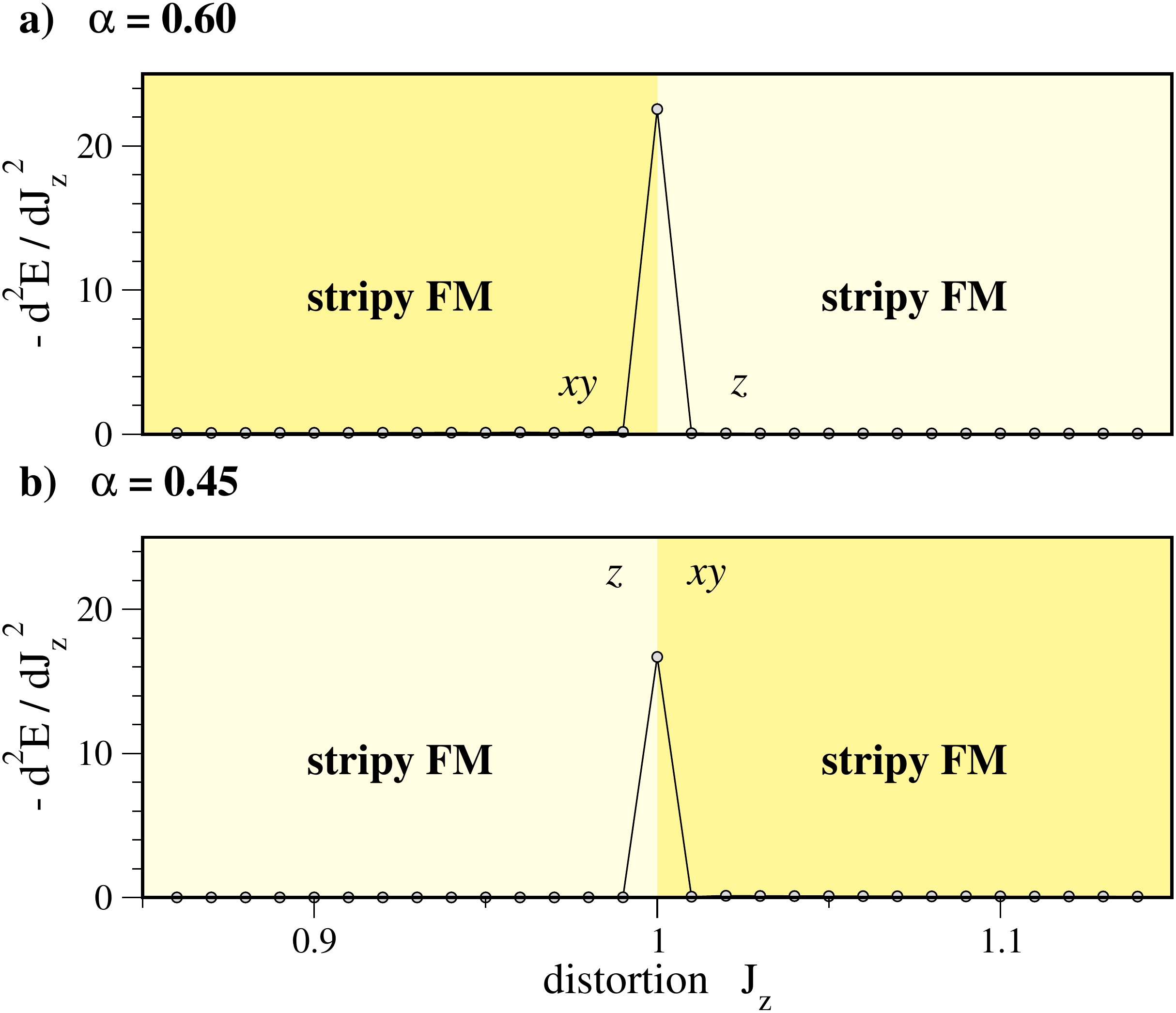}
  \caption{(Color online) Second derivative of the ground state energy density as a function of the distortion $J_z$  for two different values of the coupling $\alpha$.}
  \label{fig:energderivativeyscan2}
\end{figure}

Our approach to map out the phase diagram of the quantum Heisenberg-Kitaev model as discussed above is based on various numerical techniques, in particular exact diagonalization (ED) studies and density-matrix renormalization group (DMRG)\cite{DMRG1992} calculations for small, but highly symmetric clusters with up to $N=48$ $(N=24)$ sites for the DMRG (ED) calculations, respectively.
In order to minimize finite-size effects we employed periodic boundary conditions and chose the clusters such that they preserve the SU(2) symmetry of the four-sublattice basis transformation introduced in Section \ref{pdclassical}. We used clusters of $N=24=3\times4\times2$ and $N=32=4\times4\times2$ sites.
For the DMRG calculations we typically kept up to $m=2048$ states in the DMRG block and performed multiple sweeps to converge the observables with the typical truncation error becoming of the order of $5\times 10^{-6}$ or smaller. The location of the phase boundaries in the phase diagram (see Fig.~\ref{fig:QuantumPhaseDiagram}) are determined by the peak position of the second derivatives of the ground state energy density, i.e. ${d^2E}/{d\alpha^2}$ and ${d^2E}/{dJ_z^2}$. A similar approach has previously been used to successfully map out the phase diagram of the (undistorted) Heisenberg-Kitaev model in a magnetic field \cite{jiang}. Data for these derivatives along representative cuts in the $(\alpha, J_z)$-parameter space are shown in Figs.~\ref{fig:energderivativeyscan} and \ref{fig:energderivativeyscan2}. A very sharp peak in the second derivative -- corresponding to a jump of the first derivative of the ground state energy density, i.e., $\frac{dE}{d\alpha}$ and $\frac{dE}{dJ_z}$ -- is taken as a signature for a first-order transition and marked by the red solid lines in the phase diagram,
while a relative shallow peak in the second derivative data is interpreted as possibly indicating continuous phase transitions.

\begin{figure}
  \includegraphics[width=\linewidth]{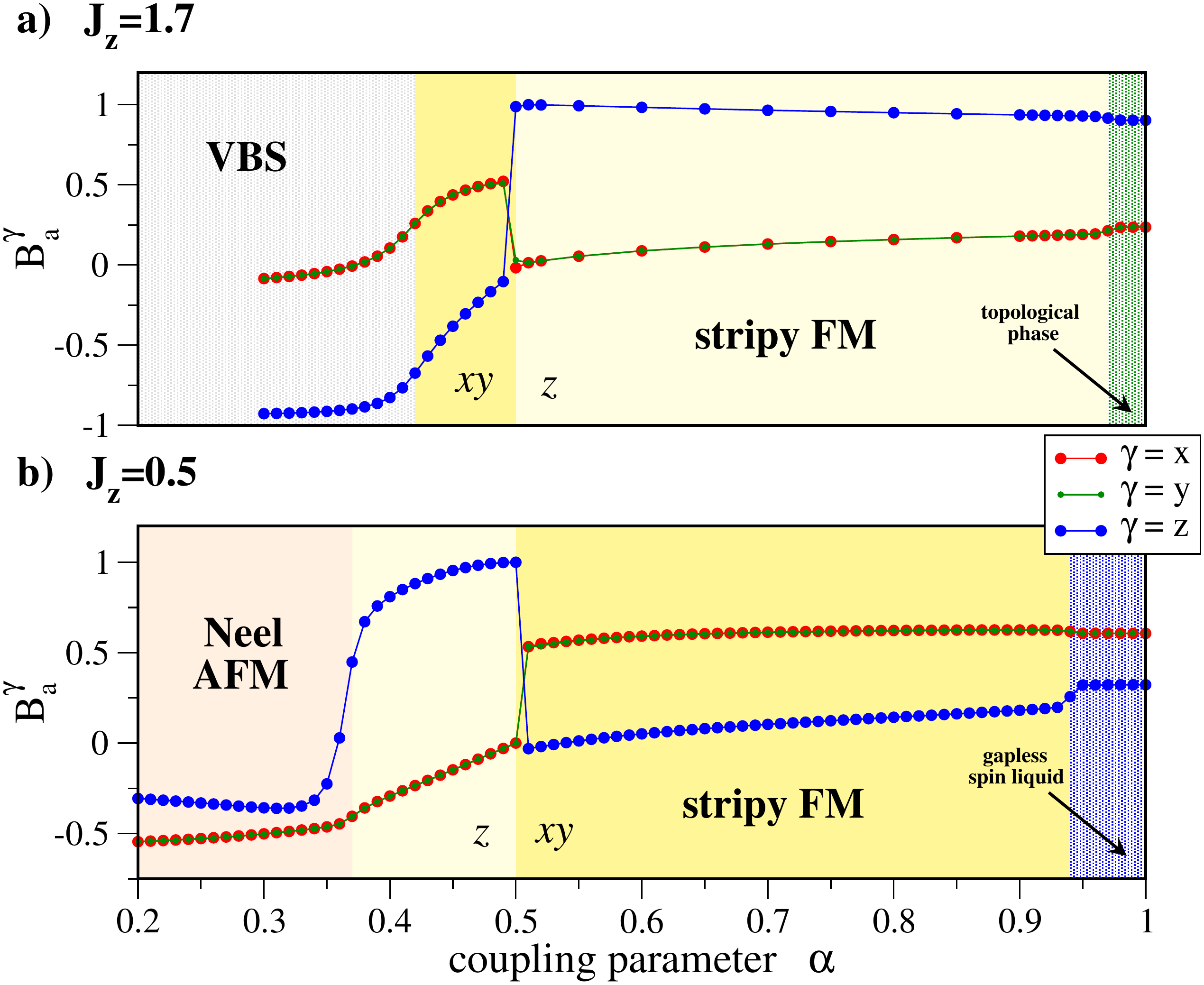}
  \caption{(Color online) Bond magnetization $B^\gamma_a$ as a function of $\alpha$ for two different values of the distortion $J_z$. Here $\gamma=a=x,y,z$.}
  \label{fig:bondoperatorscan}
\end{figure}

To further identify the nature of different phases and compare with the classical Heisenberg-Kitaev model, we calculate a `bond magnetization', i.e. the expectation value of the bond operator $B^\gamma_a=\langle S^\gamma_i\cdot S^\gamma_{i+\hat{a}}\rangle$ where $\gamma=x,y,z$ denotes the $\gamma$ component of spin, and $\hat{a}=\hat{x},\hat{y},\hat{z}$ is the unit vector along an $a$-bond.
As illustrated in Fig.~\ref{fig:bondoperatorscan} this bond magnetization is a very useful tool to track the quantum order-by-disorder selection in the distorted stripy phases. For example, in the stripy-$z$ phase for $(\alpha>1/2,J_z>1)$ and $(\alpha<1/2,J_z<1)$, the $z$-bond magnetization $B^z_z$ is positive since $S^z$ points in the same direction in $z$-bond, while $B^z_x$ and $B^z_y$ are negative because $S^z$ are antiparallel along the $x$ and $y$ bonds (not shown). In addition to the stripy phase, this bond operator can also be used to study the phase transition between different phases, which will increase or decrease rapidly across the phase boundary. As an example, we plot the bond operator $B^\gamma_a$ with $\gamma=a=x,y,z$ in Fig.~\ref{fig:bondoperatorscan}(b), in which the dotted lines are the phase boundaries determined by $B^\gamma_a$, and consistent with the ones determined by the second derivative of the ground state energy density.

Finally, we want to shortly comment on the quantum order-by-disorder mechanism playing out in the distorted stripy phase.
As mentioned earlier, for precisely $\alpha=1/2$ the system exhibits an additional $SU(2)$ symmetry and its ground state can be characterized by a conventional ferromagnetic order parameter in terms of the four-sublattice transformed $\tilde{S}$ spin variables
introduced in Section \ref{pdclassical}. For small deviations from $\alpha=1/2$ the symmetry of the model is reduced to a discrete one. But as we saw in Section~\ref{pdclassical} when discussing the classical model, one can quickly see that on the mean field level the actual direction of the ferromagnetic order is not fixed upon introducing a distortion. In fact, as we have shown in Section~\ref{pdclassical} thermal fluctuations are ultimately responsible for the eventual ordering along cubic axes in the classical model. An analogous argument applies to the quantum model where quantum fluctuations will favor a locking of the spin orientation along the cubic axes of the model for finite distortions at zero temperature. This effect has been first commented on in Refs.~\onlinecite{khaliullin} and \onlinecite{chaloupka} and is discussed in detail in appendix \ref{app:QuantumOBDO}.


\subsection{Phase transition out of the Abelian topological phase}

\begin{figure}[b]
  \begin{center}
    \includegraphics*[width=0.8\columnwidth]{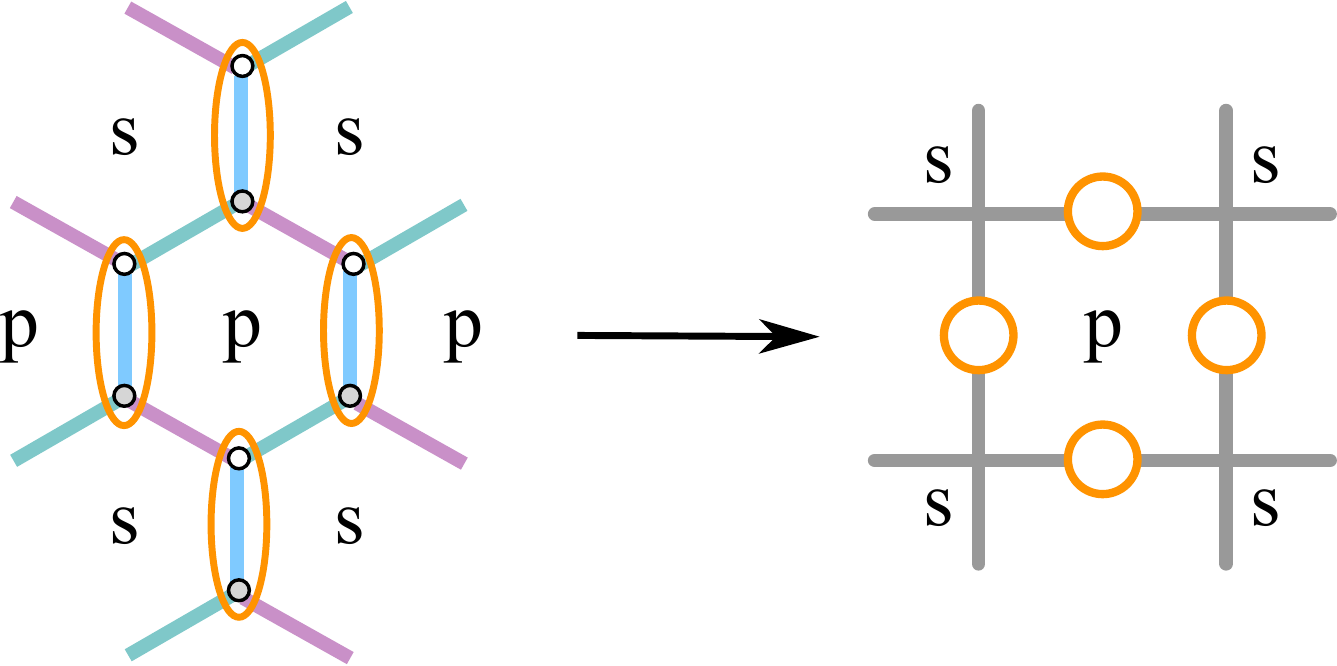}
    \caption{\label{fg:toric}Mapping from the honeycomb lattice to the toric code lattice.
    }
  \end{center}
\end{figure}

For $J_z=3$, $J_x = J_y=0$, the system decouples into $z$-dimers with Hamiltonian $H^z_{ij}$ given in Eq.~(\ref{eq:bondH}). The Heisenberg term has the singlet state $s$ as the ground state with an excited triplet $\{t^+ ,t^- ,t^0\}$, whereas the Kitaev ferromagnetic term has a degenerate pair of ground states $(t^\pm)$ and a second degenerate pair of excited states $(s,t^-)$. The energies of these states are $E_{s}/J_z =5 \alpha - 3$, $E_{t^\pm}/J_z =1-3 \alpha$, and $E_{t^0}/J_z = 1+ \alpha$.
For $\alpha=1$, one can formulate an effective interaction between the
doublet $t^\pm$ degrees of freedom localized on $z-$links and
represented by  effective spins $\sigma_i^z = \pm 1$. Thus for a given
$z-$link $ \sigma_i^z = +1$ for the state $\left |\uparrow \uparrow \right\rangle$ and $ \sigma_i^z = -1$ for the state $\left| \downarrow \downarrow \right\rangle $.
 Following Kitaev,~\cite{Kitaev} those spins can be located on the links of a square lattice; see Fig.~\ref{fg:toric}.
 For small $J_x$, $J_y$ the dimer-dimer interaction can be represented as an effective interaction between the $\sigma$'s. For the Kitaev model ($\alpha=1$) the leading interaction is generated at forth order in $J_x, J_y$, and is a 4-spin interaction equivalent to the toric code model.
Explicitly for $J_x=J_y \ll J_z$,~\cite{Kitaev}
\bea
\label{eq:toric}
H^{(4)} = -J_{\rm{TC}} \sum_P Q_P,
\eea
with $J_{TC}=\frac{J_x^2 J_y^2}{16 J_z^3}$, and the plaquette operator  $Q_P = \sigma^y_{\rm{left(p)}}   \sigma^y_{\rm{right(p)}}   \sigma^z_{\rm{up(p)}} \sigma^z_{\rm{down(p)}}$, where $P$ runs over all hexagonal plaquettes of the honeycomb lattice, which become either plaquettes $p$ or stars $s$ on the square lattice of the toric code.

In the presence of the Heisenberg term, we find an interaction already at first order in $J_{x,y}$ which reads simply
\bea
\label{eq:ising}
H^{(1)} = J \sum_{\langle i,j \rangle} \sigma_i^z  \sigma_j^z,
\eea
with $J=J_x(1-\alpha)$. Here $\langle i,j \rangle$ are nearest
neighbors in the square lattice of the toric code. One immediately
sees that this term stabilizes a N\'eel order of the effective spins
$\sigma_i^z$, which is equivalent to the stripy-$z$ phase in
Fig.~\ref{fg:stripy}. Therefore the phase transition between the
topological phase and stripy phase emanates from the right-top corner
of the phase diagram. By comparing the energy scales of the
interaction $J_{\rm{TC}}$, in Eq.~(\ref{eq:toric}), stabilizing the topological phase and the Ising interaction $J$, in Eq.~(\ref{eq:ising}), one immediately sees that the transition line approaches the right-top point as
\bea
1-\alpha \propto (3 - J_z)^3.
\eea
This high power of ($3-J_z$) is consistent with the very small area occupied by the gapped topological phase in our phase diagram in Fig.~\ref{fig:QuantumPhaseDiagram}.


\subsubsection*{Possibility of  condensation of $(e,m)$ excitations}
\begin{figure}[t]
  \begin{center}
    \includegraphics[width=.8\columnwidth]{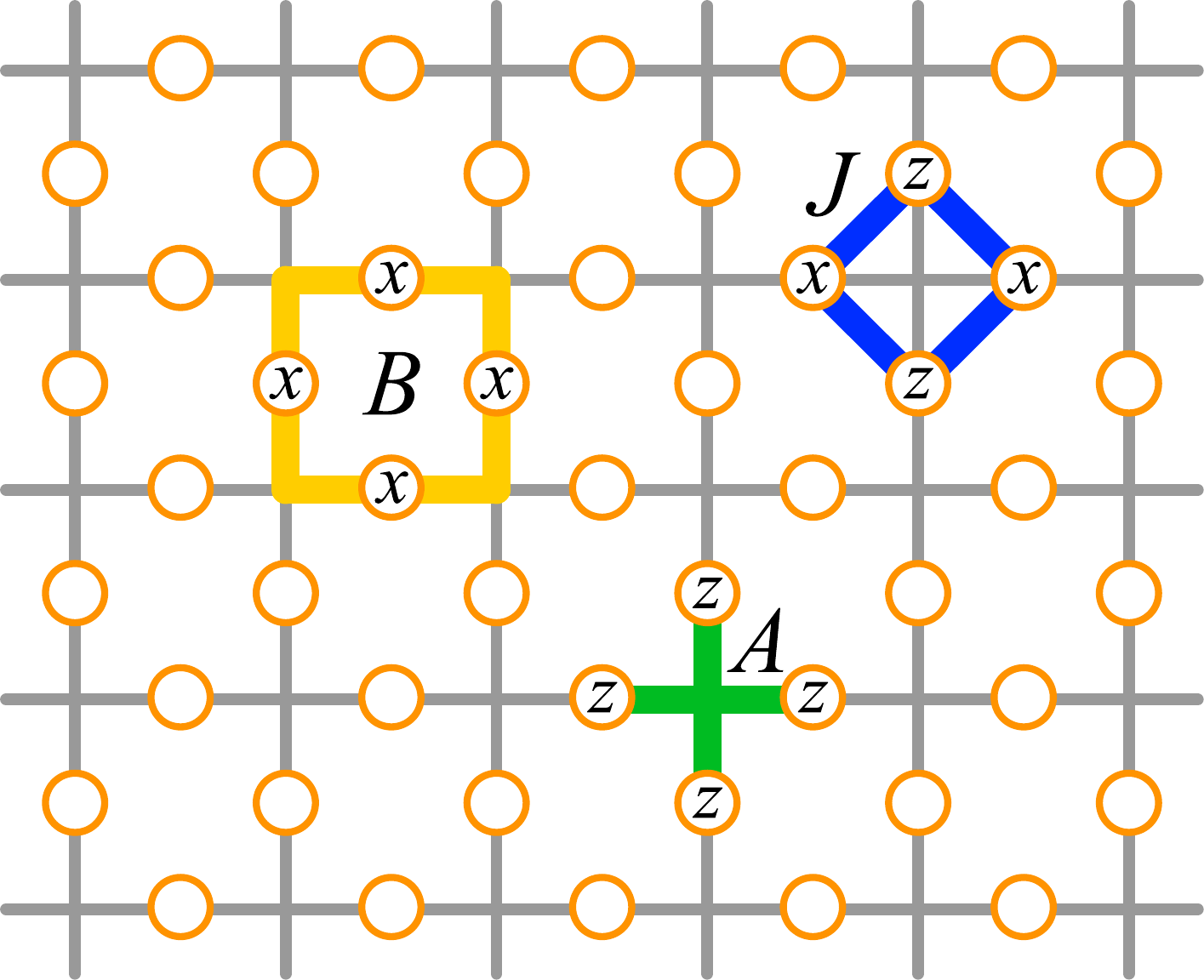}
    \caption{\label{fg:abj} $A$ represents star operators in Eq.~(\ref{eq:ABJ}), $B$ plaquette operators, and $J$ Ising coupling on nearest neighbors.
    }
  \end{center}
\end{figure}

We propose a simple model to understand the quantum phase transition between the gapped topological phase and stripy phase. This model contains just the two competing interactions which stabilize either phase: the toric code Hamiltonian Eq.~(\ref{eq:toric}) and the Ising Hamiltonian Eq.~(\ref{eq:ising}). In order to introduce standard notation for the toric code model, after permuting the spin indices $ (z,x,y) \to (x,y,z) $, and then performing a $-\pi/2$ rotation along $y$ for spins living on vertical bonds in the toric code square lattice defined in Fig.~\ref{fg:toric}, the model becomes
\bea
\label{eq:ABJ}
H = - A \sum_s \prod_{i \in s} \sigma^z_i-B \sum_p \prod_{i \in p} \sigma^x_i  + J  \sum_{\langle i ,j \rangle } \sigma^x_i \sigma^z_j,
\eea
where the coupling $J_{\rm{TC}}$ have been separated into star and plaquette operators with couplings $A$ and $B$, shown in Fig.~\ref{fg:abj}. In our case $A = B = J_{\rm{TC}}$. In the last term $\sigma^x_i$ always belongs to an horizontal bond and $\sigma^z_j$ to a vertical bond and $\langle i ,j \rangle$ are nearest neighbors; see Fig.~\ref{fg:abj}.
As a function of $J$ there must be a quantum phase transition between the $Z_2$ gapped topological phase to the Ising ordered phase at $J \sim J_{\rm{TC}}$ with spontaneously broken local Ising symmetry
\bea
\sigma^x_i &\to & - \sigma_i^x,~~~i \in ~{\rm{horizontal~link}},\nonumber \\
\sigma^z_j & \to & - \sigma^z_j,~~~j \in ~{\rm{vertical~link}}.
\eea
One can write this model in terms of the excitations of the gapped topological phase: (i) electric excitations $e$ living on stars $s$ with $-1$ eigenvalue of
\bea
e^z_s = \prod_{i \in s} \sigma^z_i,
\eea
and (ii) magnetic excitations $m$  living on plaquettes $p$ with $-1$ eigenvalue of
\bea
m^z_p = \prod_{i \in p} \sigma^x_i.
\eea
In the physical Hilbert space both $e$ and $m$ excitations occur in pairs. Such pairs are created, respectively, by
\bea
e^x_{ss'} =\prod_{i \in \mathcal{C}_{s s'}} \sigma^x_i,~~~m^x_{pp'} =\prod_{i \in \mathcal{C}_{p p'}} \sigma^z_i,
\eea
where $\mathcal{C}_{s s'}$($\mathcal{C}_{pp'}$) is an arbitrary path along the lattice (dual lattice) connecting stars $s$, $s'$ (plaquettes $p$, $p'$) where the two excitations are created.
One can check that $\{e^x_{ss'} , e^z_{s''} \}=0$ for $s''=s$ or $s''=s'$, and $[e^x_{ss'} , e^z_{s''}]=0$ otherwise, and the $m$'s satisfy similar relations. Independent of the choice of contours, $e^x_{ss'}$ and $m^x_{pp'} $ commute if the corresponding contours cross an even number of times and anticommute otherwise.

The Hamiltonian is simply
\bea
\label{eq:effectivemodel}
H(J) &=& -A \sum_s e^z_s -B  \sum_p m^z_p   \nonumber \\
&+& J \sum_{\langle i ,j \rangle }  e^x_{s_0 s_i} e^x_{s_0 s'_i} m^x_{p_0 p_j} m^x_{p_0 p'_j}.
\eea
Here each horizontal edge $i$ is shared by two stars $s_i$ and $s'_i$ and each vertical edge $j$ shares two plaquettes $p_j$ and $p'_j$. The reference star $s_0$ and reference plaquette $p_0$ are arbitrary, and can be thought of as being located at infinity (with open boundary conditions).

Clearly in the  N\'{e}el phase there is a finite expectation value of
\bea
E= \langle \sigma_i^x \rangle =\langle e^x_{s_0 s_i} e^x_{s_0 s'_i} \rangle \ne 0,
\eea
and
\bea
M=\langle \sigma_j^z \rangle = \langle m^x_{p_0 p_j} m^x_{p_0 p'_j} \rangle \ne 0,
\eea
and their relative sign is opposite for $J>0$. In the topological phase all excitations are gapped and uncorrelated. Thus a natural question is how the $e$- and $m$- excitations condense.
Typically, excitations condense at a phase transition as their kinetic energy exceeds the mass gap. From the effective model Eq.~(\ref{eq:effectivemodel}), we see that to first order in $ J$ individual $e$- and $m$- excitations can not hop thus their excitation  energy is $2A$ and $2B$, respectively. On the other hand their bound state $(e,m)$ does acquire kinetic energy of order $J$. It can hop along the $x-$direction hence lowering the gap to $2 A + 2B - 2   J$. This suggests an interesting type of quantum phase transition consisting of a condensation of the $(e,m)$ bound states for large enough $J$, which is unusual due to the fermionic nature of those composite particles. It is interesting to explore this possibility on a quantitative level in the future.


\subsection{One-dimensional limit of the Heisenberg-Kitaev model}

In the limit of $J_z=0$, corresponding to the bottom in the phase diagram in Fig.~\ref{fig:QuantumPhaseDiagram}, the system decomposes into decoupled Heisenberg-Kitaev chains. The physics of such chains has previously been partially explored, in particular with regard to its energy dynamics \cite{Steinigeweg}. Here we will apply one dimensional (1D) field theoretical methods to analytically construct the 1D phase diagram of such Heisenberg-Kitaev chains, and to gain insight into the 2D case by studying the limit of weakly coupled chains.

\subsubsection{Phase diagram}
\begin{figure}[t]
  \centering
  \includegraphics*[width=\columnwidth]{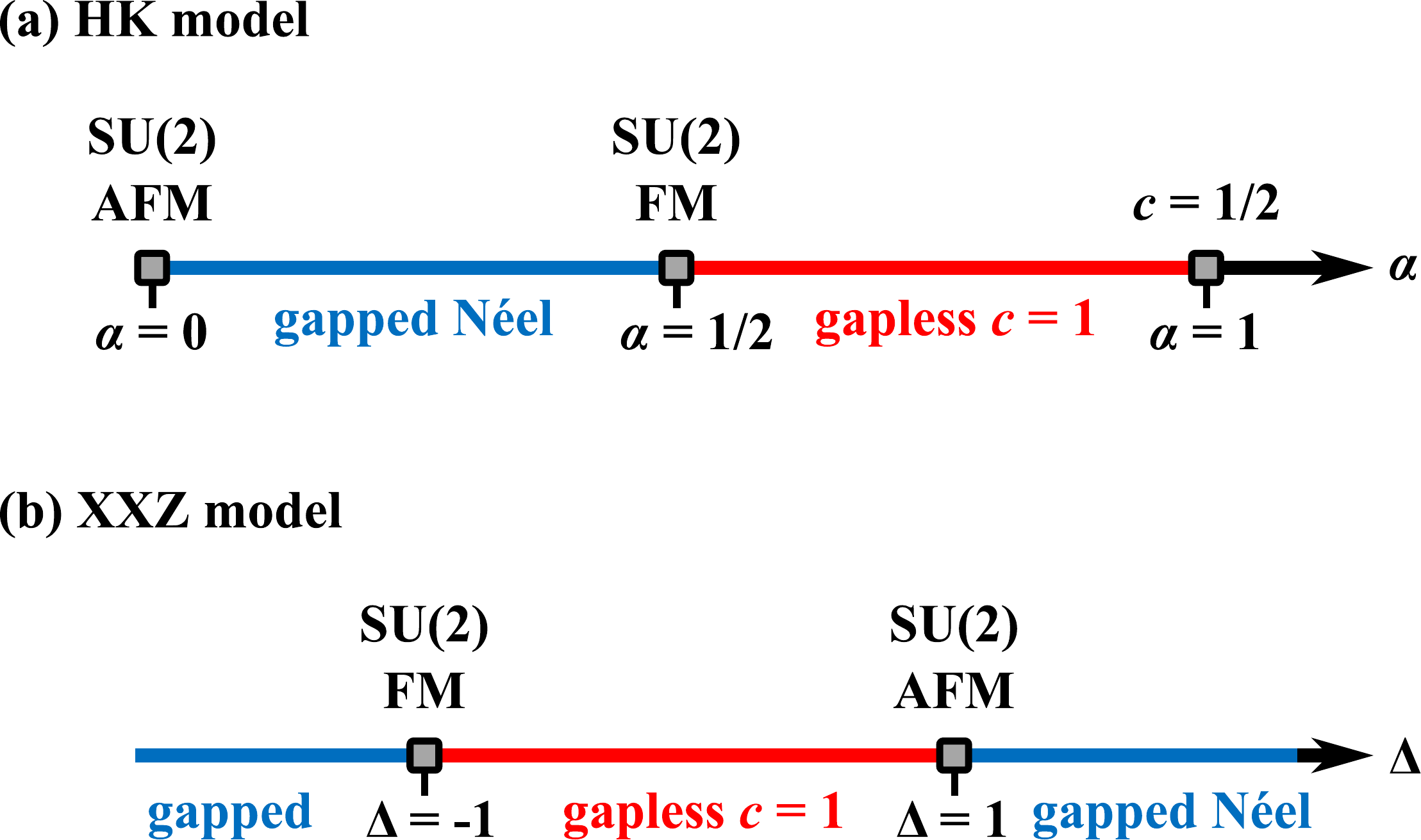}
  \caption{(a) \label{fg:1dfd}Phase diagram of the 1D Heisenberg
    Kitaev model. (b) \label{fg:xxzpd}Phase diagram of the XXZ
    model. }
\end{figure}

Our phase diagram of the 1D Heisenberg-Kitaev (HK) model is shown in Fig.~\ref{fg:1dfd}(a). It contains three exactly solvable points: (i) For $\alpha=0$ the model is the antiferromagnetic Heisenberg chain which is described by a conformal field theory (CFT) with central charge $c=1$. (ii) At $\alpha=1/2$, the model written in terms of the $\tilde{S}$ spin variables is the ferromagnetic Heisenberg chain, which has dynamical critical exponent $z=2$. (iii) At $\alpha=1$, the system is also critical and can be described by a CFT with $c=1/2$ corresponding to gapless Majorana chains~\cite{Kitaev}.
Below we describe the phases in between these three exactly solvable points.

It is convenient to express the 1D HK Hamiltonian as the sum of the well studied XXZ model,
\bea
H_{\rm{xxz}}[\Delta] = \sum_i [S^x_i S_{i+1}^x+S^y_i S_{i+1}^y+ \Delta S^z_i S_{i+1}^z],
\eea
and a perturbation. Indeed, our model reads
\bea
\frac{H_{\rm{HK}}(S)}{J_x} = (1-2 \alpha) H_{\rm{xxz}}[ \frac{1-\alpha}{1-2 \alpha} ] \pm \alpha \delta H,
\eea
with
\be
\delta H =\sum_i (-1)^{i+1} [S^x_i S_{i+1}^x-S^y_i S_{i+1}^y].
\ee
The $\pm$ signs correspond to alternating chains. The well known phase diagram of the XXZ model is summarized in Fig.~\ref{fg:xxzpd}(b).

We begin by analyzing the small $\alpha$ limit. At $\alpha=0$ the perturbation to the XXZ chain vanishes and $\Delta=1$. Our system lies inside the gapless Luttinger liquid phase of the XXZ model which extends in the range $-1 \le \Delta \le 1$. This phase is described by a Luttinger liquid theory~\cite{Giamarchi} which is characterized by Luttinger parameter $K$ and velocity $v$, given exactly by
\bea
\label{eq:Kv}
K = \frac{\pi}{2 (\pi - \arccos \Delta)},~~~
v= \frac{\pi \sqrt{1-\Delta^2}}{2 \arccos \Delta}.
\eea
We find that the perturbation to the XXZ model $\delta H$ has renormalization group scaling dimension
\bea
\label{eq:x2}
x_K = K+K^{-1}.
 \eea
Hence it is marginal $(x_K=2)$ at $K=1$ ($\Delta = 0 $ in the XXZ model), and otherwise it is irrelevant $(x_K >2)$. Since at the vicinity of the point $\alpha=0$ the perturbation $\delta H$  is both small and irrelevant, we may safely ignore it.  In other words the HK 1D model at $\alpha=0^+$ and the XXZ model at $\Delta=1^+$ differ only by the irrelevant operator $\delta H$. When $\alpha$ becomes nonzero and positive, $\Delta$ increases above unity in the XXZ chain, and then the gapless phase is destroyed and the chain undergoes a  Kosterlitz Thouless transition into a N\'eel ordered state along $z$. For the field theoretical description of this transition in the XXZ model we refer the reader to Ref.~\onlinecite{Giamarchi} and references therein.

Translating the  N\'eel order to the $\tilde{S}$ variables, one
obtains the ferromagnetic-$z$ phase, \emph{e.g.} $\uparrow \uparrow
\uparrow \uparrow$. This order parameter coincides with that of the
ferromagnetic point at $\alpha=1/2$. Therefore we expect that the
N\'eel ordered phase (in terms of the original spin variables)
persists in the entire range $0<\alpha \le 1/2$.

We now analyze the vicinity of the point $\alpha = 1/2$. It is convenient to write the HK model in terms of the $\tilde{S}$ spin variables using Eq.~(\ref{CJK}). After a $\pi-$rotation around the $z$ axis of each second spin, one can rearrange terms into a sum of an XXZ model and a perturbation,
\bea
\frac{H_{\rm{HK}}(\tilde{S}) }{J_x}= \alpha H_{\rm{xxz}}[ \frac{\alpha-1}{\alpha} ] \pm (2 \alpha - 1)\delta H.
\eea
We see that $\alpha=1/2$ brings us to the  point $\Delta=-1$ in the XXZ model. This point is connected to the gapless phase of the XXZ model, although it has different universality with vanishing velocity, see Eq.~(\ref{eq:Kv}), and dynamical critical exponent $z=2$.

We now consider $\alpha$ slightly larger than $1/2$. Since the perturbation to the XXZ model in terms of the $\tilde{S}$ spins, $\delta H$, has exactly the same form as the perturbation in terms of original spins, $S$, we draw the same conclusion regarding the irrelevance of $\delta H$. We have again a model which up to an irrelevant operator is equivalent to the XXZ model. We see that moving to the right from $\alpha=1/2$ is equivalent to moving to the right from $\Delta=-1$
 in the  XXZ model, entering into the gapless Luttinger liquid phase. We expect that the end point of the Luttinger liquid phase is the 1D limit of the  Kitaev $Z_2$ liquid, $\alpha=1$. This spin liquid does not have any continuous symmetry. This is consistent with the statement that only upon approaching the point $\alpha=1$ the operator $\delta H$ becomes relevant. Using the scaling dimension of $\delta H$, Eq.~(\ref{eq:x2}), this implies for the Luttinger liquid parameter $K \to 1$ upon approaching the Kitaev limit. Thus the region $1/2 < \alpha < 1$ maps to the region $-1<\Delta<0$ in the XXZ model.

\begin{figure}[t]
  \centering
  \includegraphics*[width=\columnwidth]{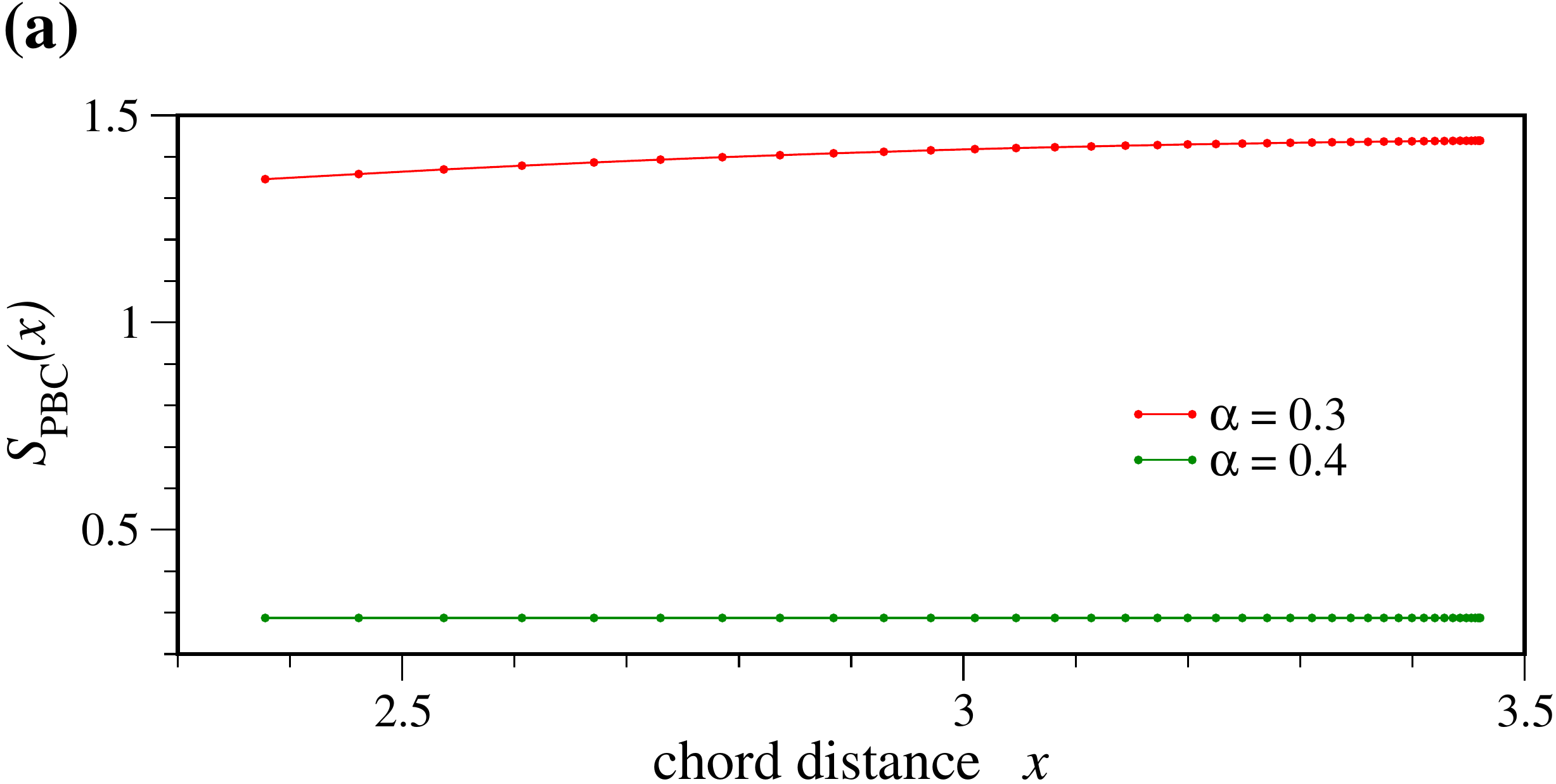}
  \includegraphics*[width=\columnwidth]{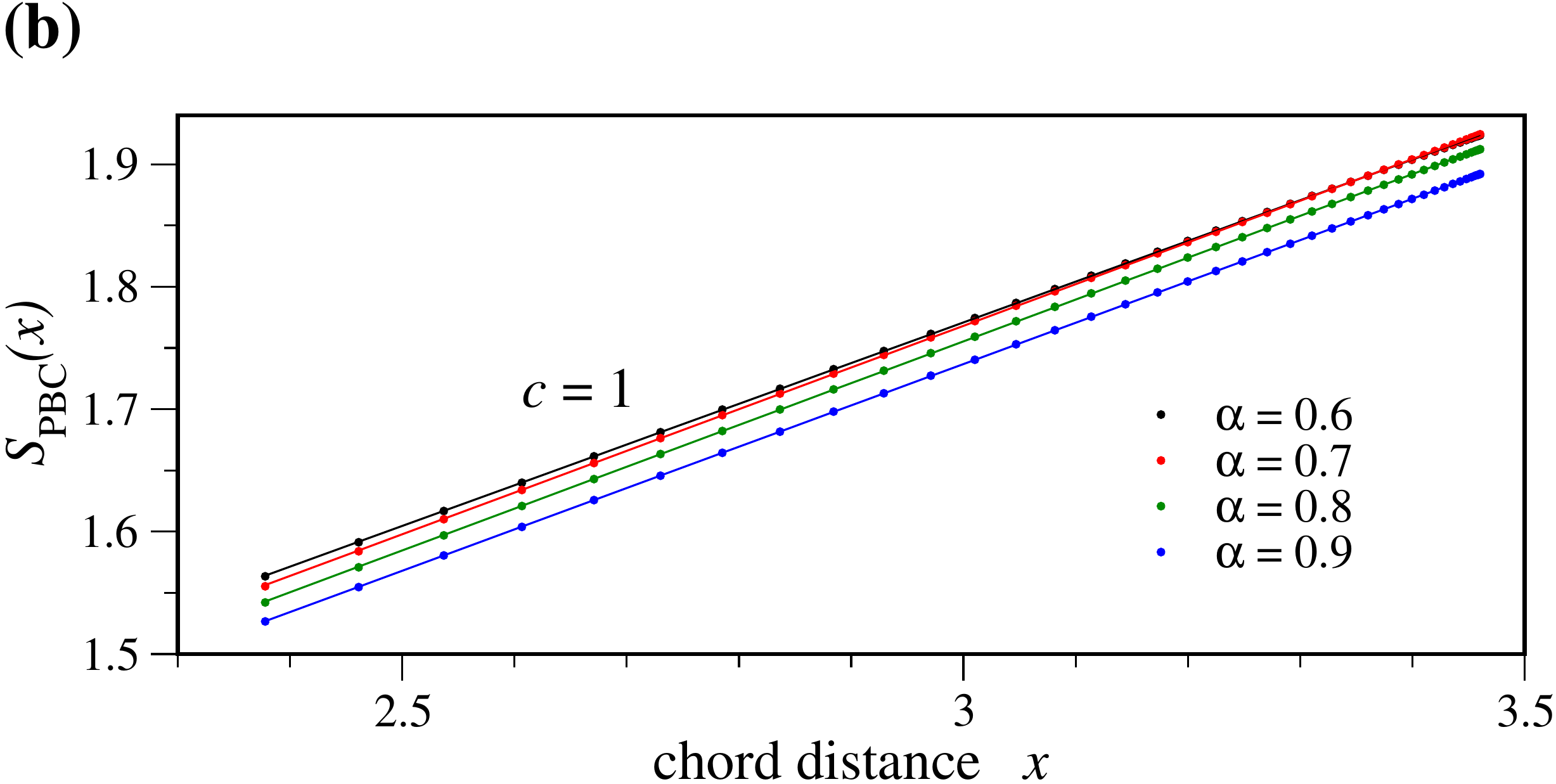}
  \includegraphics*[width=\columnwidth]{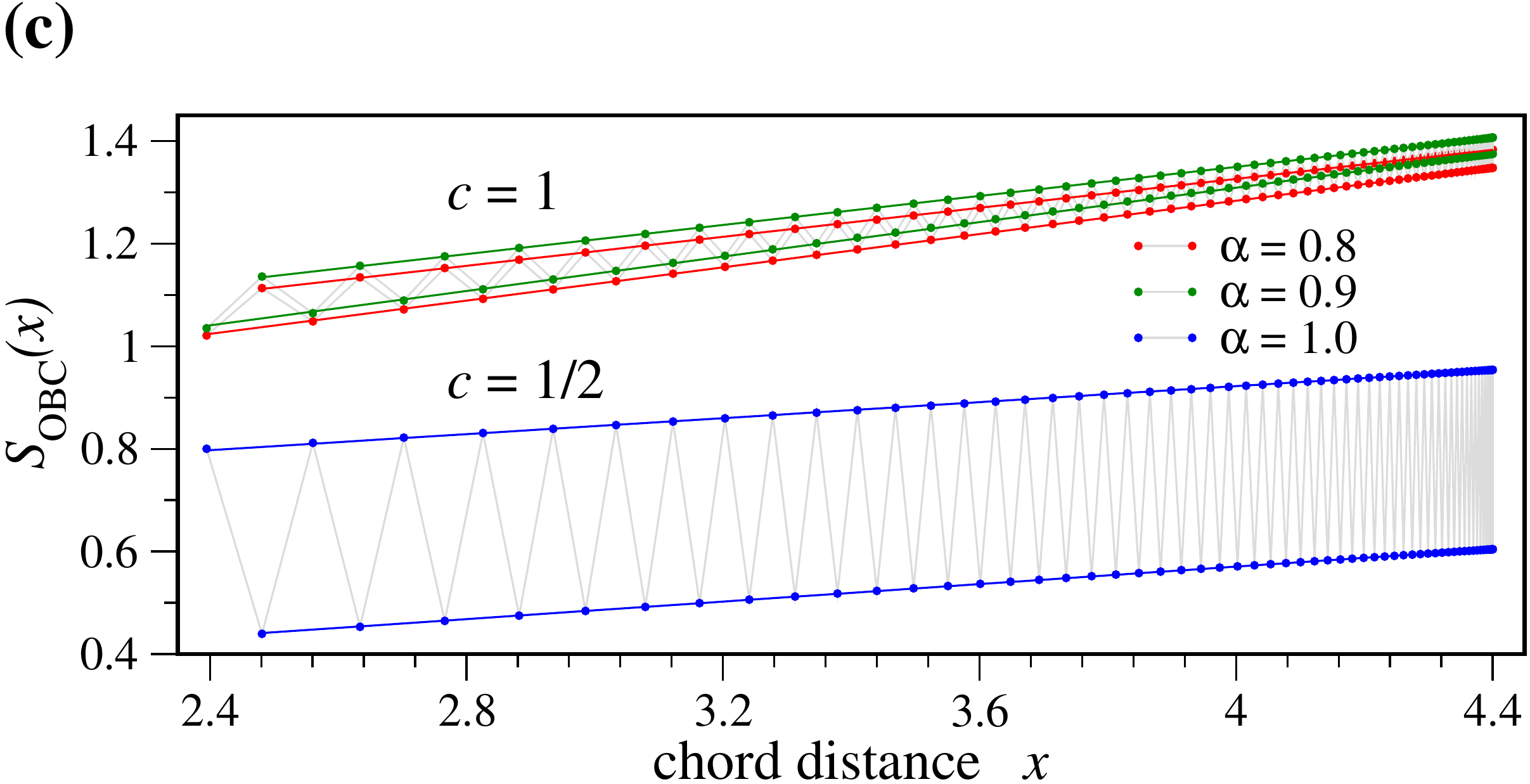}
  \caption{\label{fg:1dNumerics}
                Entanglement scaling for the 1D Heisenberg-Kitaev chain for various coupling strength $\alpha$.
                Panel (a) shows data for a periodic $L=100$ site chain in the gapped phase for $\alpha<1/2$.
                Panel (b) shows data for a periodic $L=100$ site chain in the gapless phase for $\alpha>1/2$.
                Panel (c) shows data for an open $L=256$ site chain in the gapless phase for $\alpha>1/2$ and $\alpha=1$.
                Boundary effects of the open chain result in an odd-even staggering.
                }
\end{figure}

The existence of a gapped phase for  $0 < \alpha < 1/2$ as well as the persistence of a $c=1$ gapless phase in the parameter regime $1/2 < \alpha < 1$ is nicely confirmed by  DMRG simulations of chains with open and periodic boundary conditions with up to $L=256$ sites, see Fig.~\ref{fg:1dNumerics}. For the extended gapless phase the central charge of the conformal field theory describing the gapless system can easily be extracted by fitting the entanglement entropy to the analytical form
 \begin{equation}
  S(x) = \dfrac{c}{3\eta} \cdot  x  +O(1)  \,,
  \label{eq:cft_of_l}
\end{equation}
where $x = \ln{ \left[\dfrac{\eta L}{\pi} \sin{\left( \dfrac{\pi l}{L}\right)}\right]} $ is the logarithm of the so-called chord distance for a cut dividing the chain into segments of length $l$ and $L-l$ and periodic (open) boundary conditions are indicated by the parameter $\eta=1$ or $2$, respectively.
Performing such a fit as indicated by the solid lines in Fig.~\ref{fg:1dNumerics} nicely confirms the expected central charge of $c=1$.  In the Kitaev limit $\alpha=1$ the gapless phase is verified to be described by $c=1/2$ conformal field theory, as validated by our numerical simulations~\cite{JiangBalents} shown in Fig.~\ref{fg:1dNumerics}(c).

\subsubsection{Insight about 2D}
Having constructed the phase diagram of the 1D HK model at $J_z=0$, we now consider perturbatively the coupling between the chains by studying the effect of a small $J_z$. This will be useful for the purpose of locating the precise position of the phase transition between the N\'eel and stripy phase when $J_z \to 0$, which will be the focus of this section.

In terms of the original spins the inter-chain Hamiltonian is
\bea
H_\perp = H_z &=&\sum_{{\rm{z-links}}} J_z[(1-\alpha) {\bf S}_i {\bf S}_{j} - (2 \alpha) S^z_i S^z_{j}] \nonumber \\
&=&\sum_{{\rm{z-links}}} J_z[(1-\alpha)2 (S_i^+ S_{j}^- +S_i^- S_{j}^+)\nonumber \\
&&~~~~~~~~~~+ (1-3 \alpha) S^z_i S^z_{j}].
\eea
For $0<\alpha<1/2$ the chains are ordered and gapped already on the 1D level characterized by the N\'eel order parameter $M(j)$. On the level of expectation values we have  $\langle   S^z_{i,j}  \rangle = (-1)^i  M(j)$ where $j$ labels different chains. Here the notation $ S^z_{i,j} $ refers to a deformation of the honeycomb into a brickwall lattice.  Within this ordered state the interchain coupling acts classically and couples $M(j)$ to $M(j \pm 1)$. We thus obtain two regimes for the 2D system:
\begin{enumerate}
\item{$0<\alpha<1/3$: in this regime the effective interchain coupling is antiferromagnetic, leading to $M(j)  = - M(j \pm 1)$. This phase is the 2D N\'eel antiferromagnet. The order parameter is $\sum_{i,j} (-1)^{i+j} \langle S_{i,j}^z \rangle$.}
    \item{$1/3<\alpha<1/2$: in this regime the effective interchain coupling is ferromagnetic leading to $M(j)  = M(j \pm 1)$. This phase is the stripy $z$ phase. The order parameter is $\sum_{i,j} (-1)^{i} \langle S_{i,j}^z \rangle = \sum_{i,j} \langle \tilde{S}_{i,j}^z \rangle $ corresponding to ferromagnetic order of the rotated spin variables.}
\end{enumerate}
Thus using weak chain coupling in the regime $\alpha<1/2$ we found the N\'eel  antiferromagnet as well as the stripy phase, which exist at strong interchain coupling $J_z=1$. It is therefore reasonable that the phases we found from the 1D limit are indeed connected to those found earlier along the $C_3$ symmetric line without phase transitions in between. This is indeed confirmed by our numerical calculation, see Fig.~\ref{fig:QuantumPhaseDiagram}. It is interesting that the transition between the N\'eel AF and stripy-$z$ phases along the 1D line occurs exactly at $\alpha=1/3$ as for the classical model; see Fig.~\ref{fig:distortedphasediagram}. The discrepancy with the numerical transition point at $J_z=0$ between the N\'eel AF and stripy phase could result from a finite-size effect.

%
%
\section{Summary}
\label{sec:outlook}

To conclude, using a combination of numerical and analytical methods we have established a rich phase diagram for the distorted Heisenberg-Kitaev model on the honeycomb lattice. Probably the most interesting questions left for future work concern a deeper 
understanding of the nature of the phase transition between the topological and non-topological phases in this phase diagram. 
Various recent studies~\cite{TrebstShtengel,Tupitsyn,Vidal09a,Vidal09b,Dusuel10,Dusuel11,Schmidt13,Kamfor14} have addressed the phase transitions between the gapped  $Z_2$ topological phase of Kitaev's toric code model and conventionally ordered states.  
Some of these transitions are well understood continuous phase transitions arising from the condensation of one of the elementary 
(bosonic) excitations of the toric code, often referred to as electric charges ($e$) or magnetic vortices ($m$), as it is the case for the 
phase transition induced by a single-component magnetic field pointing along one of the two longitudinal directions. 
Our analysis of the distorted Heisenberg-Kitaev model has led to an effective model, which potentially paves the path to a different
type of phase transition arising from the simultaneous condensation of the fermionic $(em)$ bound state of an electric and magnetic excitation, which drives the system from the $Z_2$ topological phase to a conventional phase with stripy order. Such fermionic $(em)$ bound states have been previously discussed in the context of the single and two-component (longitudinal) magnetic field transitions \cite{Dusuel10,Vidal09a,Tupitsyn}, transverse field transitions \cite{Vidal09b}, and in more general field theoretical terms \cite{Moon}.
Similarly, the nature of the phase transition between the stripy phase and the gapless topological phase, which has been a topic of recent interest~\cite{chaloupka,jiang,schaffer,reuther}, may be further explored in our distorted model where one can benefit from anisotropic limits.

%
%

\section{Acknowledgments}
We thank L. Balents, P. Fendley, N. Perkins,  A. Rosch, and K. P. Schmidt for  insightful discussions.
We also thank O. Wohak for his contributions to the early stages of the simulations of the classical models.
E.S. was supported by the A.V. Humboldt Foundation and an ISF grant. H.C.J. was supported by the Templeton Fund.
S.T. acknowledges hospitality of the Aspen Center for Physics and partial support from DFG SFB TR12.
The numerical simulations were performed in part on the CHEOPS cluster at RRZK Cologne.
We further acknowledge computing support from the Center for Scientific Computing at the
CNSI and MRL supported through NSF grants NSF MRSEC (DMR-1121053) and
NSF CNS-0960316, respectively.

%
%

\appendix

\section{Classical order by disorder mechanism}
\label{app:m4}
We now provide the details of the derivation of   the effective Hamiltonian Eq.~(\ref{eq:m4}) starting from the continuum model  Eq.~(\ref{cont}). This effective Hamiltonian of ${\bf \hat{e}}$ is defined via integrating over the fluctuations $\pi_a$ (defined in Eq.~(\ref{eq:pi}),
\bea
e^{- H_{\rm{eff}}[{\bf \hat{e}}]/T} =  \int \mathcal{D} \pi_a({\bf r}) e^{- \mathcal{H}[{\bf \hat{e}} ,\pi_a({\bf r})]/T}.
\eea
We now compute this effective Hamiltonian explicitly by expanding $\mathcal{H}[{\bf \hat{e}} ,\pi_a({\bf r})]$ up to quadratic order in the fluctuations $\pi_a({\bf r})$. $\mathcal{H}[{\bf \hat{e}} ,\pi_a({\bf r})]$ contains a Heisenberg part and a Kitaev part. For the Heisenberg part we use
\bea
(\nabla_{{\bf u}_\gamma} {\bf M})^2= (\nabla_{{\bf u}_\gamma} \pi_1)^2+ (\nabla_{{\bf u}_\gamma} \pi_2)^2+\mathcal{O}(\pi_a^4),
\eea
which does not depend on the magnetization direction ${\bf \hat{e}}$. For the Kitaev term we have
\bea
\nabla_{{\bf u}_\gamma} M^\gamma = \hat{e}_1^\gamma  \nabla_{{\bf u}_\gamma} \pi_1+\hat{e}_2^\gamma  \nabla_{{\bf u}_\gamma} \pi_2+\mathcal{O}(\pi^2)\,,
\eea
which depends on the magnetization direction ${\bf \hat{e}}$ through its complementary orthogonal vectors ${\bf \hat{e}}_1$ and ${\bf \hat{e}}_2$. In $k$-space,
\bea
\mathcal{H}[{\bf \hat{e}} ,\pi_a({\bf r})] =\frac{\mathcal{J}}{2 }\sum_{\bf k} \sum_{a,b} \pi_a ({\bf k}) h_{ab}({\bf k}) \pi_b (-{\bf k})\,,
\eea
with $\mathcal{J} = J/A_{\rm{hex}}$ and
\bea
h_{ab} ({\bf k})=\sum_\gamma \left( (1-\alpha)\delta_{ab}+(4 \alpha-2) e^\gamma_a e^\gamma_b \right) k_{u_\gamma}^2,
\eea
where $k_{u_\gamma} = {\bf k} \cdot {\bf \hat{u}}_\gamma$. Performing the Gaussian integrals over $\pi_a({\bf k})$ we arrive at the effective Hamiltonian
\bea
\label{eq:classicalobd}
\frac{H_{\rm{eff}}}{T} = \sum_{\bf k} \log \det \left( \frac{\mathcal{J}}{2 T } \hat{h}({\bf k}) \right),
\eea
where $\det \hat{h} = h_{11} h_{22}- h_{12} h_{21}$.
In order to proceed analytically we assume that the anisotropic Kitaev term is small, $|\alpha-1/2|\ll 1$. Then, up to a constant, we can expand
 \bea
\frac{H_{\rm{eff}}}{T} = \sum_{\bf k} \log \det
\left(\delta_{ab}+ \epsilon \sum_\gamma e^\gamma_a e^\gamma_b \frac{k_{u_\gamma}^2}{|k|^2} \right).
\eea
with small parameter $\epsilon=\frac{4}{3} \frac{2 \alpha-1}{1-\alpha} \propto (2\alpha-1)$.

We further make the simplifying approximation of a circular Brillouin zone of radius $k_{BZ}$ such that the total number of sites is $N = \sum_{\bf k} =\mathcal{A} \int_{|{\bf k}| < k_{BZ}} \frac{d ^2 k}{(2 \pi)^2}$ with total area $\mathcal{A}= N A_{\rm{hex}}$. Using polar coordinates for the momentum integral and performing the integral over $|{\bf k|}$ we obtain $\frac{H_{\rm{eff}}}{NT}  = \int_0^{2 \pi} \frac{d \theta}{2 \pi}  \log \tr (\textbf{1}+ \epsilon \hat{A})$. Next we use the identity $\log \det = \tr \log$, and expand the $\log$ up to quadratic order in $\alpha-1/2$, to obtain
\begin{equation}
\frac{H_{\rm{eff}}}{NT} = \int_0^{2 \pi} \frac{d \theta}{2 \pi}  \left[ \epsilon \tr (\hat{A}) - \frac{1}{2}\epsilon^2 \tr( A^2) \right] +\mathcal{O}(\alpha-\tfrac{1}{2})^3 ,
\end{equation}
with $A_{ab}=\sum_\gamma e^\gamma_a e^\gamma_b \cos^2(\theta - \theta_\gamma)$,  $\theta_x = 2 \pi/3$, $\theta_y = 4 \pi/3$,  $\theta_z=0$. Using $\int_0^{2 \pi} \frac{d \theta}{2 \pi} \cos^2 (\theta-\theta_\gamma) = 1/2,~~~(\gamma=x,y,z)$ we have simply for the first order term $\int_0^{2 \pi} \frac{d \theta}{2 \pi} \tr A = \frac{1}{2} \tr \sum_\gamma e_a^\gamma e_b^\gamma = \frac{1}{2}({\bf \hat{e}}_1^2+{\bf \hat{e}}_2^2)= 1$ which is a constant independent of ${\bf \hat{e}}$. The second-order term is evaluated similarly. Using $\int_0^{2 \pi} \frac{d \theta}{2 \pi} \cos^2 (\theta-\theta_\gamma) \cos^2 (\theta-\theta_{\gamma'}) = \frac{3 }{16} (1+\delta_{\gamma \gamma'})$, we have
\begin{equation}
\frac{H_{\rm{eff}}}{N T} =-\frac{2}{3} \left(2 \alpha-1 \right)^2
 \sum_{\gamma, \gamma'} \sum_{a,b=1,2}  e^\gamma_a e^\gamma_b e^{\gamma'}_b e^{\gamma'}_a (1 + \delta_{\gamma \gamma'}).
\end{equation}
As the unit vectors ${\bf \hat{e}}_1,{\bf \hat{e}}_2,{\bf \hat{e}}$ form an orthonormal basis, one can readily derive the identities $\sum_{\gamma , \gamma'} \sum_{a,b=1,2}  e^\gamma_a e^\gamma_b e^{\gamma'}_b e^{\gamma'}_a = 2$, and
\bea
\sum_\gamma \sum_{a,b=1,2}  e^\gamma_a e^\gamma_b e^\gamma_b e^\gamma_a  = \sum_\gamma (\hat{e}^\gamma)^4+\rm{const}.
\eea
As a result we obtain the decisive term in the effective Hamiltonian, up to a constant and up to quadratic order in $2 \alpha-1$,
\bea
\frac{H_{\rm{eff}}}{N T} =-\frac{2}{3 } \left( 2 \alpha-1\right)^2
\left[(\hat{e}^x)^4+(\hat{e}^y)^4+(\hat{e}^z)^4\right].
\eea


\section{Emergent magnetostatics in the classical Kitaev model}
\label{app:Coulomb}

The aim of this section is to provide a brief, self-consistent description of the Coulomb gas formulation of the spin liquid state in the undistorted classical Kitaev model, i.e. we consider the situation of $J_x = J_y = J_z = J$ only.
Given an arbitrary configuration of spins $S^\gamma$, we can assign to each lattice bond
$(\textbf{R}_i,\gamma)$ connecting sublattice $a$ site (filled circles in Fig.~\ref{fg:dimcover}) $\textbf{R}_i$ to a neighboring sublattice $b$ (empty circles in Fig.~\ref{fg:dimcover}) site $\textbf{R}_i+{\bf \hat{u}}_\gamma$,  a vector ${\bf E}=E(\textbf{R}_i;\gamma) {\bf \hat{u}}_\gamma$, with
\bea
E(\textbf{R}_i;\gamma) = (S_{a,\textbf{R}_i}^\gamma)^2 - \frac{1}{3}.
\eea
The discrete divergence of the ${\bf E}$-field at vertices $\textbf{R}_i$ of sublattice $a$ vanishes by definition, $\sum_\gamma E(\textbf{R}_i;\gamma)=0$, since $|{\bf S}_{\textbf{R}_i}|=1$. The nontrivial property of the ground states of the classical Kitaev model is that they satisfy a divergence-free condition also in the $b$ sublattice,
\bea
{\rm{vertex}}~ \textbf{R}_i+{\bf \hat{u}}_z:~~~{\boldsymbol \nabla} \cdot {\bf E} = \sum_\gamma E(\textbf{R}_i - \textbf{r}_\gamma;\gamma)=0,
\eea
where $\textbf{r}_\gamma = {\bf \hat{u}}_z - {\bf \hat{u}}_\gamma$. This condition follows from the formation of dimer-covering states; see Fig.~\ref{fg:dimcover}. In such states for every spin on sublattice $a$ there exists a neighboring spin in sublattice $b$ such that both spins point ferromagnetically along the direction of the connecting bond.

It is not difficult to show that the dimer covering states have the lowest possible energy for the classical Kitaev model.\cite{chandra} The partition function
\bea
Z = \int \prod_{\textbf{R}_i} \frac{d S_{a,\textbf{R}_i}}{4 \pi} \frac{d S_{b,\textbf{R}_i}}{4 \pi}e^{- H/T}
\eea
can be evaluated by writing the Hamiltonian as $H =- \sum_{\textbf{R}_i} {\bf S}_{a,\textbf{R}_i} {\bf B}_{\textbf{R}_i}$, where ${\bf B}_{\textbf{R}_i} = \sum_\gamma J_\gamma S^\gamma_{b,\textbf{R}_i+{\bf \hat{u}}_\gamma} {\bf \hat{u}}_{\gamma}$, and then performing the integral over spins of sublattice $a$ which appear to be free except for an external field ${\bf B}_{\textbf{R}_i}$. This gives
\bea
Z = \int \prod_{\textbf{R}_i} \frac{d S_{b,\textbf{R}_i}}{4 \pi}e^{-  \sum_{\textbf{R}_i} h_{{\rm{eff}}}[|{\bf B}_{\textbf{R}_i}|]/T}.
\eea
Now using the convexity of the effective Hamiltonian $h_{{\rm{eff}}}(B) = -T \log \frac{T \sinh ( B/T)}{ B}$, which implies $\langle h_{{\rm{eff}}}(x) \rangle < h_{{\rm{eff}}}(\langle x \rangle)$, one sees that the total energy is minimized when all $|{\bf B}_{\textbf{R}_i}|$'s are equal. As $\sum_{\textbf{R}_i} |{\bf B}_{\textbf{R}_i}|=N J$ the minimum occurs when $|{\bf B}_{\textbf{R}_i}|=J$. This situation is indeed achieved in the dimer-covering state. It should be noted that there exist an additional continuous slide degree of freedom within the ground state.~\cite{shankar}

The emergent divergence-free $\bf E$-field leads to peculiar features in observables that depend on $\bf E$. For example consider the bond-energy correlation $\langle (S_i^z)^2 (S_j^z)^2 \rangle-1/9$ which measures the $\langle E^z E^z \rangle$ correlation. The following simple derivation applies to Coulomb phases in general, so we now coarse grain the original lattice and consider separations $|i-j|$ much larger than the lattice spacing.
For such a long distance description we can think of ${\bf E}$ as a field on a continuous space which satisfies the divergence free condition
\bea
\label{eq:div0}
{\boldsymbol \nabla} \cdot {\bf E}({\bf r})=0.
\eea
At zero temperature all divergence free field configurations are equally likely. At finite temperature (low enough to avoid considerable charge density) field configurations having locally a net polarization are suppressed entropically, leading to the leading quadratic term in the effective free energy
\bea
F = \frac{\mathcal{K}}{2} \int d^d r [{\bf E}({\bf r})]^2.
\eea
Here the constant $\mathcal{K}$ is analogous to the permittivity in electrodynamics and this coarse grain formulation may be considered in arbitrary dimension $d$.
The divergence-free constraint Eq.~(\ref{eq:div0}) is easily taken into account in momentum space where it reads
\bea
\textbf{k} \cdot {\bf E}(\textbf{k})=0,
\eea
and the free energy is $F = \frac{\mathcal{K}}{2} \sum_{\textbf{k}}  |E^\perp (\textbf{k}) |^2$. Here $E^\perp (\textbf{k})$ refers to the components (single component in 2D) of ${\bf E}$ perpendicular to $\textbf{k}$. The correlation function is calculated directly from the equipartition $
\langle E^\mu(-\textbf{k}) E^\nu(\textbf{k}') \rangle =\mathcal{K}^{-1} P_{\mu \nu} \delta_{\textbf{k} ,\textbf{k}'}$, where in a basis whose first element is parallel to $\textbf{k}$ we have $P=\left(
                                                 \begin{array}{cc}
                                                   0 & 0 \\
                                                   0 & \textbf{1}_{d-1} \\
                                                 \end{array}
                                               \right)
$. Writing the projector $P$ in a general basis gives
\bea
\langle E^\mu(-\textbf{k}) E^\nu(\textbf{k}') \rangle =\mathcal{K}^{-1} \left( \delta_{\mu \nu}-\frac{k_\mu k_\nu}{|\textbf{k}|^2} \right) \delta_{\textbf{k} ,\textbf{k}'}.
\eea
This implies a power-law decay of correlation functions in real space $\langle E^\mu(-\textbf{k}) E^\nu(\textbf{k}') \rangle \propto \frac{1}{|{\bf r}|^d}$.
 This general result implies pinch points in correlation functions, since the correlation functions depend on how the limit $|{\bf k}| \to 0$  is approached. For our model this implies for the $(S^z_i)^2$ correlation the form
\bea
\label{eq:pinch}
S(\textbf{k}) = \langle (S_i^z)^2 (S_j^z)^2 \rangle_{\textbf{k}} \propto \frac{k_y^2}{k_x^2 + k_y^2},
\eea
leading to the pinch point at $|{\bf k}|=0$.


\section{Quantum order-by-disorder mechanism} 
\label{app:QuantumOBDO}

Following a similar logic as in the classical case, here we will determine the magnetization direction of the quantum stripy phase, by following an order-by-disorder calculation of the fluctuations. Technically the quantum fluctuations will be taken into account by a large$-S$ expansion of the $\tilde{S}$ spin variables, with Hamiltonian~(\ref{CJK}).
We represent the spins in sublattice $a$ and $b$ using Holstein-Primakoff bosons $a_{R_i}$ and $b_{R_i}$, respectively, as
\bea
{\bf \tilde{S}}_{a,R_i} &=& (S - a^\dagger_{R_i} a_{R_i}) {\bf \hat{e}} \nonumber \\
&+&\sqrt{\frac{S}{2} }  [{\bf \hat{e}}_1  (a_{R_i}+a_{R_i}^\dagger)-i {\bf \hat{e}}_2  ( a_{R_i}-a_{R_i}^\dagger )], \nonumber \\
{\bf \tilde{S}}_{b,R_i} &=& (S - b^\dagger_{R_i} b_{R_i}) {\bf \hat{e}} \nonumber \\
&+&\sqrt{\frac{S}{2} }  [{\bf \hat{e}}_1  (b_{R_i}+b_{R_i}^\dagger)-i {\bf \hat{e}}_2  ( b_{R_i}-b_{R_i}^\dagger )].
\eea
We have expanded around a uniform ground state with magnetization direction $\bf{\hat{e}}_1 \times \bf{\hat{e}}_2 = {\bf \hat{e}}$. The terms in the Hamiltonian of order $S^2$ give the classical energy. At this level the energy is independent on the magnetization direction ${\bf \hat{e}}$. We now evaluate the next leading order terms in a $1/S$ expansion. It is convenient to compute the spin-spin couplings appearing in the Hamiltonian. For the Heisenberg we have the simple form
\bea
{\bf \tilde{S}}_{a} \cdot {\bf \tilde{S}}_{b} = -S(a^\dagger a +b^\dagger b - a^\dagger b -a b^\dagger)
\eea
For the Kitaev term we have
\bea
{\bf \tilde{S}}^\gamma_{a}  {\bf \tilde{S}}^\gamma_{b} = \nonumber \\
\left( (S - a^\dagger  a){\bf \hat{e}}^\gamma +\sqrt{\frac{S}{2} } [ {\bf \hat{e}}_1^\gamma (a+a^\dagger) -i {\bf \hat{e}}_2^\gamma  ( a-a^\dagger )] \right) \times \nonumber \\
\left( (S - b^\dagger  b){\bf \hat{e}}^\gamma+\sqrt{\frac{S}{2} } [ {\bf \hat{e}}_1^\gamma (b+b^\dagger) -i {\bf \hat{e}}_2^\gamma  ( b-b^\dagger )] \right)
\eea
This includes an $\mathcal{O}(S^{3/2})$ term linear in the bosons $a$, $a^\dagger$, $b$, and $b^\dagger$. This linear term contains one contribution proportional to ${\bf \hat{e}}^\gamma {\bf \hat{e}}_1^\gamma$, and another proportional to ${\bf \hat{e}}^\gamma {\bf \hat{e}}_2^\gamma$. Upon summing over the three links connected to either $a$ or $b$ these contributions vanish since ${\bf \hat{e}} \cdot {\bf \hat{e}}_1={\bf \hat{e}} \cdot {\bf \hat{e}}_2=0$.

The $\mathcal{O}(S)$ term is quadratic in the bosonic operators. After Fourier transformation the $\mathcal{O}(S)$ term can be written as
\bea
\label{eq:c4}
\frac{\tilde{H}}{S} = \sum_{k} [\Psi_k^\dagger \hat{h}_k \Psi_k+(2-4 \alpha) \sum_\gamma J_\gamma ({\bf \hat{e}}^\gamma)^2],\nonumber \\
\Psi_k = \left(
  \begin{array}{cccc}
    a_k & b_k & a^\dagger_{-k} & b^\dagger_{-k} \\
  \end{array}
\right) ^T \eea
with
\widetext
\bea
\hat{h}_k &=&\sum_\gamma J_\gamma [\frac{\alpha-1}{2}\left(
  \begin{array}{cccc}
    -1 & e^{i {\bf k} \cdot r_\gamma} & 0 & 0 \\
    e^{-i {\bf k} \cdot r_\gamma} & -1 & 0 & 0 \\
    0 & 0 & -1 & e^{i {\bf k} \cdot r_\gamma} \\
    0 & 0 & e^{-i {\bf k} \cdot r_\gamma} & -1 \\
  \end{array}
\right) \nonumber \\&+& \frac{1-2\alpha}{2}
\left(
  \begin{array}{cccc}
    -2 ({\bf \hat{e}}^\gamma)^2 & (({\bf \hat{e}}^\gamma_1)^2+({\bf \hat{e}}^\gamma_2)^2) e^{i {\bf k} \cdot r_\gamma} & 0 & ({\bf \hat{e}}^\gamma_1+i {\bf \hat{e}}^\gamma_2)^2 e^{i {\bf k} \cdot r_\gamma} \\
    (({\bf \hat{e}}^\gamma_1)^2+({\bf \hat{e}}^\gamma_2)^2) e^{-i {\bf k} \cdot r_\gamma} & -2 ({\bf \hat{e}}^\gamma)^2 & ({\bf \hat{e}}^\gamma_1+i {\bf \hat{e}}^\gamma_2)^2 e^{-i {\bf k} \cdot r_\gamma} & 0 \\
    0 & ({\bf \hat{e}}^\gamma_1-i {\bf \hat{e}}^\gamma_2)^2 e^{i {\bf k} \cdot r_\gamma} &  -2 ({\bf \hat{e}}^\gamma)^2  & (({\bf \hat{e}}^\gamma_1)^2+({\bf \hat{e}}^\gamma_2)^2) e^{i {\bf k} \cdot r_\gamma} \\
     ({\bf \hat{e}}^\gamma_1-i {\bf \hat{e}}^\gamma_2)^2 e^{-i {\bf k} \cdot r_\gamma}& 0 & (({\bf \hat{e}}^\gamma_1)^2+({\bf \hat{e}}^\gamma_2)^2) e^{-i {\bf k} \cdot r_\gamma} &  -2 ({\bf \hat{e}}^\gamma)^2 \\
  \end{array}
\right)]. \nonumber
\eea
\endwidetext

\begin{figure}
  \begin{center}
    \includegraphics*[width=80mm]{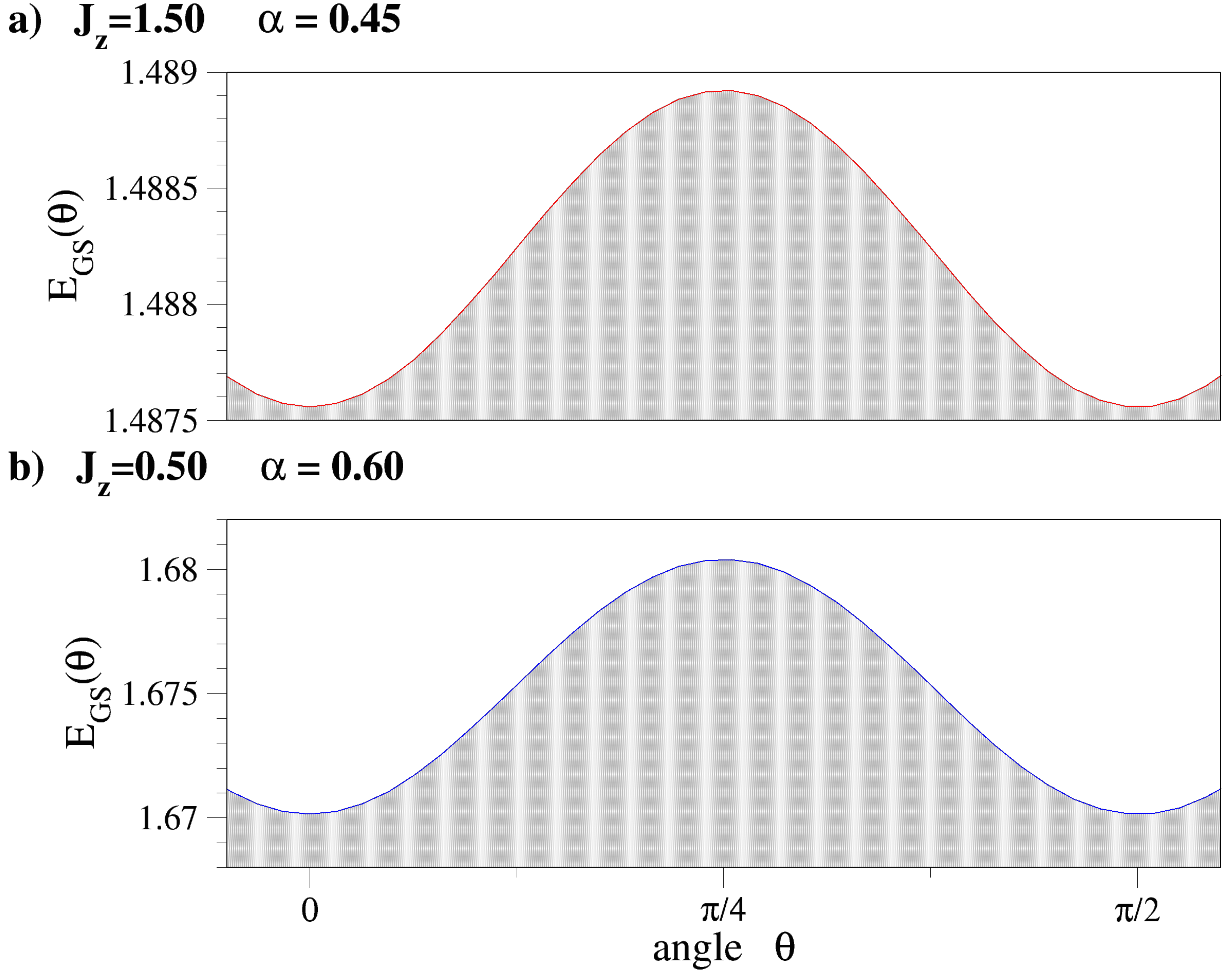}
    \caption{\label{fg:egs1}Quantum fluctuation contribution to the ground state energy per unit cell as function of azimuthal angle along the equator as parameterizing the magnetization ${\bf \hat{e}}$, with $\alpha=0.45$ $J_z=1.5$, or alternatively for $\alpha=0.6$ $J_z=0.5$ showing minima along cubic axes. Numerical evaluation of the $k$-sum in Eq.~(\ref{egs}) has been done for a honey comb lattice with $10 \times 10$ unit cells with periodic boundary conditions.
    }
  \end{center}
\end{figure}

The second term in Eq.~(\ref{eq:c4}) is proportional to the classical term $\propto S^2$. Hence it does not lift the degeneracy of the ground state manifold. Ignoring this term, after Bogoliubov transformation~\cite{sachdev92}
the quadratic Hamiltonian becomes \bea \sum_{k}  \sum_{\mu=1}^2 \omega_{k, \mu} (\Gamma^\dagger_{k \mu} \Gamma_{k \mu}+\Gamma_{k \mu} \Gamma^\dagger_{k \mu} ), \eea where $\omega_{k ,\mu} >0$ ($\mu=1,2$), and $(\omega_{k,1},\omega_{k,2},-\omega_{k,1},-\omega_{k,2})$ are the eigenvalues of the matrix $ {\rm{diag}} \{ 1,1,-1,-1\} \cdot h_k $ and $[\Gamma_{k \mu} ,\Gamma^\dagger_{k' \nu} ]=\delta_{\mu \nu} \delta_{k   k'} $. Finally, this calculation gives the $\mathcal{O}(S)$ zero point energy fluctuations per site as a function of ${\bf \hat{e}}$,
\bea
\label{egs}
E_{GS} [{\bf \hat{e}}] = \frac{1}{N}\sum_k \sum_{\mu=1}^2 \omega_{k,\mu}.
\eea
This is the quantum analog of Eq.~(\ref{eq:classicalobd}). Evaluating $E_{GS}$ numerically we find that it is minimized for ${\bf \hat{e}}$ parallel to the cubic axes for either $\alpha>1/2$ or $\alpha<1/2$ (and at $\alpha=1/2$). For example $E_{GS}$ is plotted for $\alpha=0.45$ and $J_z=1.5$, or alternatively for  $\alpha=0.6$ and $J_z=0.5$ in Fig~(\ref{fg:egs1}) demonstrating that the magnetization points along the cubic axes in the stripy $xy$ phases.

Whereas our linear spin wave calculation gives also the dispersion of
the spin-waves, $\omega_{k,\mu}$, it fails to show the opening of the
gap of the Goldstone modes once the continuous symmetry is spoiled at
$\alpha \ne 0$. A self-consistent spin-wave calculation does account
for the gap in the spin-wave spectrum.\cite{khaliullin,chaloupka}

%
%

\input{DistortedHKModel.bbl}

\end{document}

%% file: DistortedHKModel.bbl
%